\documentclass[prb,superscriptaddress,showpacs,floatfix,tightenlines,10pt,twocolumn]{revtex4-1}
\usepackage{amssymb}
\usepackage{amsmath}
\usepackage{graphicx}
\usepackage{float}
\usepackage{xcolor}

\newcommand{\func}[1]{\operatorname{#1}}

\begin{document}

\title{Internodal excitonic state in a Weyl semimetal in a strong magnetic
field}
\author{Ren\'{e} C\^{o}t\'{e}}
\author{Gautier D. Duchesne}
\author{Santiago F. Lopez}
\date{\today }

\begin{abstract}
The simplest Weyl semimetal (WSM) with broken time-reversal symmetry
consists of a pair of Weyl nodes located at wave vectors $\mathbf{K}_{\tau
}=\tau \mathbf{b}$ in momentum space with $\tau =\pm 1$ the node index and
chirality. The electronic dispersion in a small wave vector region near each
node is linear and isotropic. In a magnetic field $\mathbf{B}=B\widehat{%
\mathbf{z}}$, this band structure is modified into a series of positive and
negative energy Landau levels $n=\pm 1,\pm 2,\ldots ,$which disperse along
the direction of the magnetic field, and a chiral Landau level, $n=0$, with
a linear dispersion given by $e_{\tau ,n=0}\left( k_{z}\right) =-\tau
\hslash v_{F}k_{z},$ where $k_{z}$ is the component of the electron wave
vector $\mathbf{k}$ along the direction of the magnetic field and $v_{F}$ is
the Fermi velocity. In the extreme quantum limit and for a small doping, the
Fermi level is in the chiral levels near the Dirac point. It has been shown
before that, when Coulomb interaction is considered, a Weyl semimetal may be
unstable towards the formation of a condensate of internodal electron-hole
pairs which gives rise in real space to an excitonic charge density wave.
This new state of matter is usually studied by using a short-range
interaction between the electrons. In this article we use the full
long-range Coulomb interaction and the self-consistent Hartree-Fock
approximation to generate the condensed state. We study its stability with
respect to a change in the Fermi velocity, doping and strength of the
Coulomb interaction and also consider the situation where the Weyl nodes
have a higher Chern number $C=2,3$ and more complex excitonic states are
possible. We derive the response functions and collective excitations of the
excitonic state working in the generalized random-phase approximation
(GRPA). We show that, in the mean-field gap induced by the internodal
coherence, there is, in the excitonic response function, a series of bound
electron-hole states (excitons) with a binding energy that decreases until
the renormalized Hartree-Fock energy gap is reached. In addition, there is a
collective mode gapped at exactly the plasmon frequency. By contrast, the
plasmon mode is the only excitation present in the density and current
response functions. Despite the U(1) symmetry of the excitonic state, there
is no gapless mode in the GRPA excitonic response. Indeed, the gapless mode
present in the proper excitonic response function is pushed to the plasmon
frequency by the long-range Coulomb interaction.
\end{abstract}

\maketitle

\affiliation{D\'{e}partement de physique and Institut Quantique, Universit\'{e} de
Sherbrooke, Sherbrooke, Qu\'{e}bec, Canada J1K 2R1 }

\section{INTRODUCTION}

The simplest model of a Weyl semimetal (WSM) with broken time-reversal
symmetry consists of two Weyl nodes with opposite topological charges $C=\pm
1$ located in the Brillouin zone at the wave vectors $\mathbf{K}_{\tau
}=\tau \mathbf{b,}$ where $\tau =\pm 1$ is the node index. Near each node,
the dispersion is linear and isotropic \textit{i.e.} $E_{s}=s\hslash
v_{F}\left\vert \mathbf{k}\right\vert ,$ where the wave vector $\mathbf{k}$
is measured with respect to the Weyl points and $s=\pm 1$ is the band index%
\cite{Review}. In the presence of an external magnetic field $\mathbf{B}$
directed along the $z$ axis, the band structure is transformed into a set of
positive ($n>1$) and negative ($n<1$) Landau levels that disperse along the
direction of the magnetic field according to $e_{\tau ,n\neq 0}\left(
k_{z}\right) =\frac{\hslash v_{F}}{\ell }$sgn$\left( n\right) \sqrt{%
k_{z}^{2}\ell ^{2}+2\left\vert n\right\vert },$ where $k_{z}$ is the
component of the wave vector $\mathbf{k}$ parallel to the magnetic field, $%
v_{F}$ is the Fermi velocity and $\ell =\sqrt{\hslash /eB}$ is the magnetic
length. In addition, there is a single chiral $n=0$ Landau level at each
node that disperses linearly along the direction of the magnetic field
according to $e_{\tau ,n=0}\left( k_{z}\right) =-\tau \hslash v_{F}k_{z}.$

In the extreme quantum limit at zero temperature and in the absence of
doping, the Fermi level is at the Dirac point in the chiral levels. It has
been shown\cite{Yang2011} that, in this situation, the internodal Coulomb
exchange interaction can couple electrons and holes with different
chiralities. This spontaneously hybridizes the two nodes of opposite
chiralities and opens a gap in the chiral levels. The resulting state from
this chiral symmetry breaking is an internodal condensate of electron-hole
pairs that is characterized by a complex order parameter of the form $%
\left\langle \rho _{-,+}\right\rangle =(1/N_{\varphi
})\sum_{k_{z},X}\left\langle c_{k_{z},X,-}^{\dagger
}c_{k_{z},X,+}\right\rangle =\left\vert \left\langle \rho
_{-,+}\right\rangle \right\vert e^{i\varphi },$ where $X$ is the
guiding-center index in the Landau gauge, $N_{\varphi }=S/2\pi \ell ^{2}$ is
the Landau level degeneracy (with $S$ the area of the WSM\ perpendicular to
the magnetic field), the operators $c_{k_{z},X,\tau }\left( c_{k_{z},X,\tau
}^{\dagger }\right) $ destroys(creates) an electron in state $k_{z},X,\tau $
and $\left\langle \cdots \right\rangle $ denotes a ground-state average.
These electron-hole pairs are loosely called excitons although they are not
bound states but electrons and holes paired by the internodal exchange
interaction and then condensed. The energy of this excitonic state is
independent of the phase $\varphi $ of its order parameter and so one would
expect a Goldstone mode to be associated with this U(1) symmetry.

This excitonic state has been extensively studied in the literature (see for
example, Refs. \onlinecite{Yang2011}-\onlinecite{Mottola2024}). In real
space, the excitonic condensate leads to the formation of a charge density
wave (CDW) with density $\left\langle n\left( z\right) \right\rangle \sim
\left\vert \left\langle \rho _{-,+}\right\rangle \right\vert \cos \left(
2b_{z}z+\varphi \right) $ and so to nonlinear transport properties. The
sliding motion of this incommensurate CDW (the phason), after depinning from
the impurities, is the Goldstone mode associated with fluctuations in the
phase $\varphi $. Fluctuations in the amplitude of the order parameter are
expected to be gapped. One important property of the excitonic CDW state is
that its coupling with the electromagnetic field leads to an extra
magnetoelectric axionic term in the action. The CDW is thus an example of an
axionic state of matter\cite{Roy2015} with the phason being an axion, a
hypothetical particle first proposed in high-energy physics\cite%
{Quin1977,Wilczek}. The formation of the excitonic condensate can occur with
or without the presence of a magnetic field. Excitonic states with more
complex order parameters are possible when $B=0$ involving either internodal
or intranodal electron-hole pairings.

In this paper, we study the spontaneous internodal excitonic state that can
occur between the two chiral $n=0$ Landau levels in the strong magnetic
field limit where the Fermi level lies near the Dirac point and the upper
Landau levels are empty. Contrary to most previous papers where the gap
equation for the excitonic state is derived using a contact interaction, we
use the full long-range Coulomb interaction. We solve the self-consistent
Hartree-Fock equations for the single particle Green's function numerically
using an iterative method. We study how the internodal coherence $%
\left\langle \rho _{-,+}\right\rangle =\left\langle \rho _{+,-}\right\rangle
^{\ast }$ depend on the Fermi velocity, Chern number $C=1,2,3$, doping and
strength of the Coulomb interaction using realistic values for these
parameters. For higher Chern numbers $C=2,3,$ there are respectively two and
three degenerate chiral levels at each node and more complex excitonic
states are possible involving internodal and/or interlevel coherences. We
find that a change in the Fermi velocity $v_{F}$ can produce a phase
transition between two different excitonic states.

We also study the response functions and collective excitations in the
excitonic state for the specific case of $C=1$. We derive these responses in
the generalized random-phase approximation (GRPA). As far as we know, this
has not been made before for the excitonic phase of a WSM in the strong
magnetic field limit. A similar calculation has been done for a Dirac
semimetal\cite{Macdo2017} but the band structure was different and
internodal coherence was not considered. For $C=1,$ one can define sixteen
basic response functions and the internodal coherence couples them all so
that they have to be calculated numerically. From these sixteen response
functions, we obtain the density, current and excitonic responses (they are
defined in Sec. VI). Because of a cancellation between self-energy and
vertex corrections, the only collective excitation in the current $\chi
_{jj} $ and density $\chi _{nn}$ responses is the plasmon whose frequency is
slightly modified, in the excitonic state, from its known value\cite{Plasmon}
$\omega _{p}\left( Q\right) =\sqrt{\frac{e^{3}v_{F}B}{2\pi ^{2}\varepsilon
_{0}\hslash ^{2}}+v_{F}^{2}Q^{2}}$ in an incoherent (normal)\ state. Because
of the linear dispersion of the chiral states, there is no continuum of
electron-hole excitations in these two responses as it is transformed into
the plasmon mode.

A more interesting function is the excitonic response $\chi _{\text{exc}}$
since it contains excitations related to the fluctuations in the amplitude
and phase of the complex order parameter $\left\langle \rho
_{-,+}\right\rangle $. When only the ladder diagrams are considered in the
GRPA \textit{(i.e.} the "proper" response), we find in $\chi _{\text{exc}}$
a series of electron-hole bound states (excitons) with binding energy $%
e_{B,n}$ where $n=1,2,3,...$. The energy of these resonances increases until
a continuum of electron-hole scaterring states is reached at an energy $E_{%
\text{conti}},$ which is the Hartree-Fock energy gap red-shifted by the
vertex corrections. The ladder diagrams in $\chi _{\text{exc}}\left( \omega
,Q\right) $ and the coupling between the different response functions caused
by the internodal coherence produce a gapless collective mode with frequency 
$\omega =v_{F}Q.$ When the bubble diagrams (the long-range Coulomb
interaction) are considered in the calculation,\textit{\ i.e.} within the
full GRPA, the binding energy of the excitonic states are only slightly
modified but the gapless mode is transformed into the gapped plasmon mode
present also in the density and current responses. There is no remaining
gapless collective excitation in the GRPA\ collective mode spectrum of the
excitonic state. Similar results are obtained when the GRPA is applied to
the study of collective excitations in superconductors\cite%
{Anderson,Bardasis}.

This paper is organized as follows:\ in Sec. II, we present the Hartree-Fock
description of the excitonic phase for a WSM where only the chiral levels in
each node are considered. Numerical results for this phase are given in Sec.
III for Chern number $C=1$ and in Sec. IV for $C=2,3.$ The GRPA\ approach
for the response functions in the coherent phase is described in Sec. V. We
define the current, density, and excitonic response functions $\chi
_{jj},\chi _{nn}$ and $\chi _{\text{exc}}$ in Sec. VI. Exact analytical
results are obtained for $\chi _{jj}$ and $\chi _{nn}$ but $\chi _{\text{exc}%
}$ has to be calculated numerically. Our numerical results for these
response functions are presented in Secs. VII\ and VIII for the incoherent
and coherent phases respectively. We conclude in Sec. IX. The general
Hartree-Fock formalism for Weyl nodes with an arbitrary number of Landau
levels is described in the appendix A. The precise form of the exchange
interactions that intervene in the Hartree-Fock formalism for Chern numbers $%
C=1,2,3$ are listed in the appendix B.

\section{HARTREE-FOCK DESCRIPTION OF THE EXCITONIC PHASE\qquad}

We consider a simple model of a Weyl semimetal (WSM) with broken
time-reversal symmetry consisting of two nodes, with Chern number $C,$
centered at wave vectors $\mathbf{b}=-\tau b\widehat{\mathbf{z}}$ and with
opposite chiralities $\tau =\pm 1.$The noninteracting Hamiltonian for each
node, written in the basis of the two bands that cross, is given in the
absence of a magnetic field by%
\begin{equation}
h_{\tau }\left( \mathbf{k}\right) =\tau \hslash v_{F}\left( 
\begin{array}{cc}
k_{z} & \beta \left( k_{x}-ik_{y}\right) ^{C} \\ 
\beta \left( k_{x}+ik_{y}\right) ^{C} & -k_{z}%
\end{array}%
\right) ,
\end{equation}%
where $v_{F}$ is the Fermi velocity, $\mathbf{k}$ is a wave vector measured
from the position of each Weyl node in momentum space, $\beta $ is a
material dependent anisotropy factor and $C=1,2,3$ is the Chern number.

The derivation of the Landau levels in a magnetic field $\mathbf{B}=\mathbf{%
\nabla }\times \mathbf{A=}B\widehat{\mathbf{z}}$ and of the Hartree-Fock
Hamiltonian and equation of motion for the single particle Green's function
is given in appendix A. We present there the general case where an arbitrary
number of Landau levels are kept and all types of coherence are considered
(inter-Landau-level, internodal and complete entanglement). In this section
we adapt these results to the simplest case where only the $C$ chiral levels
in each node are kept in the Hilbert space. These levels are degenerate and
have the dispersion%
\begin{equation}
e_{\tau }\left( k\right) =-\tau \hslash v_{F}k_{z}.
\end{equation}%
The corresponding eigenvectors are independent of $\tau $ and given by%
\begin{equation}
w_{n,k_{z},X}\left( \mathbf{r}_{\bot },z\right) =\frac{1}{\sqrt{L_{z}}}%
e^{ik_{z}z}\left( 
\begin{array}{c}
0 \\ 
h_{n,X}\left( \mathbf{r}_{\bot }\right)%
\end{array}%
\right) ,
\end{equation}%
with the integer $n$ taking the values $n=0$ to $n=C-1.$ In the Landau
gauge, with the vector potential $\mathbf{A}=\left( 0,Bx,0\right) ,$ the
wave functions of the two-dimensional electron gas are given by $%
h_{n,X}\left( \mathbf{r}_{\bot }\right) =\varphi _{n}\left( x-X\right)
e^{-iXy/\ell ^{2}}/\sqrt{L_{y}}$, where $X$ is the guiding-center index and $%
\varphi _{n}\left( x\right) $ the wave functions of the one-dimensional
harmonic oscillator. Each state $\left( n,k_{z},\tau \right) $ has
degeneracy $N_{\varphi }=S/2\pi \ell ^{2},$ where $S=L_{x}L_{y}$ is the area
of the WSM\ perpendicular to the magnetic field, $\ell =\sqrt{\hslash /eB}$
is the magnetic length and $\mathbf{r}_{\bot }$ is a two-dimensional vector
in the plane perpendicular to the magnetic field. The dimensions of the WSM\
are $L_{x}\times L_{y}\times L_{z}.$ Since we keep only the chiral levels,
we approximate the electron field operator by%
\begin{equation}
\Psi _{\tau }\left( \mathbf{r}\right) \approx
\sum_{n,k_{z},X}w_{n,k_{z},X}\left( \mathbf{r}_{\bot }\right)
c_{n,k_{z},X,\tau },  \label{fieldop}
\end{equation}%
where $c_{n,k_{z},X,\tau }$ annihilates an electron in state $\left(
n,k_{z},X,\tau \right) .$ To simplify the notation, we will write $k$
instead of $k_{z}$ hereafter.

For the many-body Hamiltonian of the electron gas, we take%
\begin{eqnarray}
H &=&\sum_{\tau }\int d^{3}r\Psi _{\tau }^{\dagger }\left( \mathbf{r}\right)
h_{\tau }\left( \mathbf{r}\right) \Psi _{\tau }\left( \mathbf{r}\right) 
\notag \\
&&+\frac{1}{2}\sum_{\tau ,\tau ^{\prime }}\int d^{3}r\int d^{3}r^{\prime
}\Psi _{\tau }^{\dagger }\left( \mathbf{r}\right) \Psi _{\tau ^{\prime
}}^{\dagger }\left( \mathbf{r}^{\prime }\right)  \notag \\
&&\times V\left( \mathbf{r}-\mathbf{r}^{\prime }\right) \Psi _{\tau ^{\prime
}}\left( \mathbf{r}^{\prime }\right) \Psi _{\tau }\left( \mathbf{r}\right) ,
\end{eqnarray}%
where the long-range Coulomb interaction is given by ($\varepsilon _{r}$ is
the relative dielectric constant of the WSM) 
\begin{equation}
V\left( \mathbf{r}\right) =\frac{1}{V}\sum_{\mathbf{q}}\frac{e^{2}}{%
\varepsilon _{r}\varepsilon _{0}\left\vert q_{\bot
}^{2}+q_{z}^{2}\right\vert }e^{i\mathbf{q}_{\bot }\cdot \mathbf{r}_{\bot
}}e^{iq_{z}z}.  \label{coulomb}
\end{equation}%
We have kept in $H$ only one combination of field operators that conserves
the number of electrons at each node. A second but weaker combination is
discussed in appendix A.

To fully characterize a particular phase of the electron gas in the WSM, we
use the set of ground-state averages $\left\{ \left\langle \rho
_{n,n^{\prime }}^{\left( \tau ,\tau ^{\prime }\right) }\left( k\right)
\right\rangle \right\} $ where the operators%
\begin{equation}
\rho _{n,n^{\prime }}^{\left( \tau ,\tau ^{\prime }\right) }\left( k\right) =%
\frac{1}{N_{\varphi }}\sum_{X}c_{n,k,X,\tau }^{\dagger }c_{n^{\prime
},k,X,\tau ^{\prime }}.
\end{equation}%
In terms of these operators, the Hartree-Fock Hamiltonian, for a phase that
is not modulated spatially, is given by

\begin{eqnarray}
H_{HF} &=&N_{\varphi }\sum_{n,k,\tau }e_{\tau }\left( k\right) \rho
_{n,n}^{\left( \tau ,\tau \right) }\left( k\right)  \notag \\
&&-\frac{N_{\varphi }}{L_{z}}\sum_{\tau ,\tau ^{\prime
}}\sum_{k_{1},k_{2}}\sum_{n_{1},...,n_{4}}X_{n_{1},n_{2},n_{3},n_{4}}\left(
k_{2}-k_{1}\right)  \notag \\
&&\times \left\langle \rho _{n_{1},n_{4}}^{\left( \tau ,\tau ^{\prime
}\right) }\left( k_{1}\right) \right\rangle \rho _{n_{3},n_{2}}^{\left( \tau
^{\prime },\tau \right) }\left( k_{2}\right)
\end{eqnarray}%
where the Fock interactions $X_{n_{1},n_{2},n_{3},n_{4}}\left( k\right) $
are defined in Eq. (\ref{a4}). They are nonzero for $%
n_{1}-n_{2}+n_{3}-n_{4}=0$ only so that there are $1,6,$ and $19$ nonzero
interactions of which $1,4,$ and $10$ are different for $C=1,2,3$
respectively. The Hartree term is absent from $H_{HF}$ since it is canceled
by the positive background of the WSM.

The diagonal components $\left\langle \rho _{n,n}^{\left( \tau ,\tau \right)
}\left( k\right) \right\rangle $ give the occupation $\nu \in \left[ 0,1%
\right] $ of the state $\left( n,k,\tau \right) .$ The operator $%
c_{nX,k,\tau }^{\dagger }c_{n^{\prime }X,k,-\tau }$ creates an electron-hole
pairing between the states $n,X,k,\tau $ and $n^{\prime },X,k,-\tau ,$ so
that a nonzero value of $\left\langle \rho _{n,n}^{\left( \tau ,-\tau
\right) }\left( k\right) \right\rangle $ signals a condensate of internodal
electron-hole pairs in Landau level $n,$ while $\left\langle \rho
_{n,n^{\prime }\neq n}^{\left( \tau ,\tau \right) }\left( k\right)
\right\rangle $ signals a condensate of inter-Landau-level electron hole
pairs in node $\tau $. The general case $\left\langle \rho _{n,n^{\prime
}}^{\left( \tau ,-\tau \right) }\left( k\right) \right\rangle $ with $n\neq
n^{\prime }$ represents a full entanglement between the paired electron and
hole. We loosely speak of these pairs as "excitons" although they are not
bound states. We use the words excitonic state or coherent state to refer to
the state where some type of coherence is nonzero. As we show below, such
states are favored by the exchange part (the Fock pairing in $H_{HF}$) of
the Coulomb interaction.

We explain in the appendix how the $\left\langle \rho _{n,n^{\prime
}}^{\left( \tau ,\tau ^{\prime }\right) }\left( k\right) \right\rangle
^{\prime }$s are obtained by solving the equation of motion for the
single-particle Matsubara Green's function

\begin{equation}
G_{n,n^{\prime }}^{\left( \tau ,\tau ^{\prime }\right) }\left( k,\tau
\right) =-\frac{1}{N_{\phi }}\sum_{X}\left\langle T_{\tau _{0}}c_{n,k,X,\tau
}\left( \tau _{0}\right) c_{n^{\prime },k,X,\tau ^{\prime }}^{\dagger
}\left( 0\right) \right\rangle ,
\end{equation}%
where $T_{\tau _{0}}$ is the imaginary time ordering operator and $\tau _{0}$
in the parenthesis is the imaginary time (not to be confused with the node
index). When $\tau _{0}=0^{-},$ we have

\begin{eqnarray}
\left\langle \rho _{n^{\prime },n}^{\left( \tau ^{\prime },\tau \right)
}\left( k\right) \right\rangle &=&G_{n,n^{\prime }}^{\left( \tau ,\tau
^{\prime }\right) }\left( k,\tau _{0}=0^{-}\right)  \notag \\
&=&\frac{1}{\beta \hslash }\sum_{i\omega _{n}}e^{-i\omega
_{n}0^{-}}G_{n,n^{\prime }}^{\left( \tau ,\tau ^{\prime }\right) }\left(
k,i\omega _{n}\right) ,
\end{eqnarray}%
where the fermionic Matsubara frequencies $\omega _{n}$ are defined by $%
\omega _{n}=\left( 2n+1\right) \pi /\beta \hslash $ with $n=0,\pm 1,\pm
2,... $ and where $\beta =1/k_{B}T.$ with $T$ the temperature and $k_{B}$
the Boltzmann constant.

The equation of motion is given by%
\begin{gather}
\left[ i\omega _{n}-\frac{1}{\hslash }\left( e_{\tau }\left( k\right) -\mu
\right) \right] G_{n,n^{\prime }}^{\left( \tau ,\tau ^{\prime }\right)
}\left( k,i\omega _{n}\right)  \label{motion} \\
-\frac{1}{\hslash }\sum_{\tau ^{\prime \prime },n^{\prime \prime }}\Sigma
_{n,n^{\prime \prime }}^{\left( \tau ,\tau ^{\prime \prime }\right) }\left(
k\right) G_{n^{\prime \prime },n^{\prime }}^{\left( \tau ^{\prime \prime
},\tau ^{\prime }\right) }\left( k,i\omega _{n}\right) =\delta _{\tau ,\tau
^{\prime }}\delta _{n,n^{\prime }},  \notag
\end{gather}%
where the Fock self-energies are defined as%
\begin{eqnarray}
\Sigma _{n,n^{\prime }}^{\left( \tau ,\tau ^{\prime }\right) }\left(
k\right) &=&-\frac{1}{L_{z}}\sum_{k_{1}}\sum_{n_{1},n_{2}}X_{n_{1},n^{\prime
},n,n_{2}}\left( k-k_{1}\right)  \notag \\
&&\times \left\langle \rho _{n_{1},n_{2}}^{\left( \tau ^{\prime },\tau
\right) }\left( k_{1}\right) \right\rangle
\end{eqnarray}%
and $\mu $ is the chemical potential. Equation (\ref{motion}) is solved in
the manner explained in the appendix. The ground-state energy per volume, $%
E_{HF},$ is then given by%
\begin{eqnarray}
E_{HF} &=&\frac{1}{2\pi \ell ^{2}L_{z}}\sum_{n,k,\tau }e_{\tau }\left(
k\right) \left\langle \rho _{n,n}^{\left( \tau ,\tau \right) }\left(
k\right) \right\rangle  \notag \\
&&-\frac{1}{4\pi \ell ^{2}L_{z}^{2}}\sum_{\tau ,\tau ^{\prime
}}\sum_{k_{1},k_{2}}\sum_{n_{1},...,n_{4}}X_{n_{1},n_{2},n_{3},n_{4}}\left(
k_{2}-k_{1}\right)  \notag \\
&&\times \left\langle \rho _{n_{1},n_{4}}^{\left( \tau ,\tau ^{\prime
}\right) }\left( k_{1}\right) \right\rangle \left\langle \rho
_{n_{3},n_{2}}^{\left( \tau ^{\prime },\tau \right) }\left( k_{2}\right)
\right\rangle .
\end{eqnarray}

When $C=1,$ there are then only two states to consider: $\left( n=0,\tau
=+\right) $ and $\left( n=0,\tau =-\right) $ and we can obtain some
manageable analytical results. Dropping the $n=0$ index, we write in this
case the Green's function matrix as%
\begin{equation}
G\left( k,i\omega _{n}\right) =\left( 
\begin{array}{cc}
G_{+,+}\left( k,i\omega _{n}\right) & G_{+,-}\left( k,i\omega _{n}\right) \\ 
G_{-,+}\left( k,i\omega _{n}\right) & G_{-,-}\left( k,i\omega _{n}\right)%
\end{array}%
\right) .
\end{equation}%
It satisfies the equation of motion (in matrix form)%
\begin{equation}
\left[ I\left( i\hslash \omega _{n}+\mu \right) -F\left( k\right) \right]
G\left( k,i\omega _{n}\right) =\hslash I_{2\times 2},
\end{equation}%
where $I_{2\times 2}$ is the $2\times 2$ unit matrix and the matrix $F\left(
k\right) $ is defined by%
\begin{equation}
F\left( k\right) =\left( 
\begin{array}{cc}
e_{+}\left( k\right) +\Sigma _{+,+}\left( k\right) & \Sigma _{+,-}\left(
k\right) \\ 
\Sigma _{-,+}\left( k\right) & e_{-}\left( k\right) +\Sigma _{-,-}\left(
k\right)%
\end{array}%
\right) .  \label{debut}
\end{equation}%
The self-energies are given by ($i,j=\pm $) 
\begin{equation}
\Sigma _{i,j}\left( k\right) =-\frac{1}{L_{z}}\sum_{k_{1}}X\left(
k-k_{1}\right) \left\langle \rho _{j,i}\left( k_{1}\right) \right\rangle
\label{sigma}
\end{equation}%
and the interaction%
\begin{eqnarray}
X\left( x\right) &=&X_{0,0,0,0}\left( x\right)  \notag \\
&=&\frac{e^{2}}{2\pi \varepsilon _{0}\varepsilon _{r}}\frac{1}{2}\Gamma
\left( 0,\frac{x^{2}}{2}\right) e^{\frac{x^{2}}{2}},
\end{eqnarray}%
where $\Gamma \left( 0,x\right) $ is the incomplete Gamma function. To avoid
the divergence at $x=k=0$ of this interaction, we add a very small screening
parameter $\eta $ to the Coulomb interaction in Eq. (\ref{coulomb}) \textit{%
i.e.} $y^{2}\rightarrow y^{2}+\eta ^{2}$.

The hermitian matrix $F\left( k\right) $ can be diagonalized for each value
of $k$ by solving the equation 
\begin{equation}
F\left( k\right) U\left( k\right) =U\left( k\right) D\left( k\right) ,
\end{equation}%
where $U\left( k\right) $ is the matrix of the eigenvectors and $D\left(
k\right) $ the diagonal matrix of the eigenvalues. The Green's function is
then obtained from%
\begin{equation}
G_{i,j}\left( k,i\omega _{n}\right) =\sum_{a=\pm }\frac{U_{i,a}\left(
k\right) U_{a,j}^{\dag }\left( k\right) }{i\omega _{n}-\left( E_{a}\left(
k\right) -\mu \right) /\hslash }
\end{equation}%
and, at $T=0$ K, the order parameters are given by 
\begin{equation}
\left\langle \rho _{j,i}\left( k\right) \right\rangle =\sum_{a=\pm
}U_{i,a}\left( k\right) U_{a,j}^{-1}\left( k\right) \Theta \left(
e_{F}-E_{a}\left( k\right) \right) ,  \label{fin}
\end{equation}%
where $e_{F}$ is the Fermi level which can be positive (electron doping) or
negative (hole doping) and $\Theta \left( x\right) $ is the Heaviside
function. We consider that the doping, if present, is the same in both
nodes. Equations (\ref{debut})-(\ref{fin}) constitute a self-consistent
system of equations that must be solved numerically.

The band structure in the coherent phase consists of two bands with
dispersion 
\begin{equation}
E_{\pm }\left( k\right) =\frac{1}{2}\left( \Sigma _{+}\left( k\right)
+\Sigma _{-}\left( k\right) \right) \pm \frac{1}{2}\zeta \left( k\right) ,
\label{bandes}
\end{equation}%
where%
\begin{equation}
\zeta \left( k\right) =\sqrt{\left( 2e\left( k\right) -\Sigma _{+}\left(
k\right) +\Sigma _{-}\left( k\right) \right) ^{2}+4\left\vert \Sigma \left(
k\right) \right\vert ^{2}}
\end{equation}%
and%
\begin{equation}
e\left( k\right) =\hslash v_{F}k.
\end{equation}%
The analytical expressions for the order parameters is, for each wave vector 
$k,$ 
\begin{eqnarray}
\left\langle \rho _{\pm ,\pm }\right\rangle &=&\mp \frac{1}{2\zeta }\left(
\Sigma _{+}-\Sigma _{-}-2e\mp \zeta \right) \Theta \left( e_{F}-E_{-}\right)
\notag \\
&&\pm \frac{1}{2\zeta }\left( \Sigma _{+}-\Sigma _{-}-2e\pm \zeta \right)
\Theta \left( e_{F}-E_{+}\right) ,  \label{eq12} \\
\left\langle \rho _{-,+}\right\rangle &=&-\frac{\Sigma }{\zeta }\left[
\Theta \left( e_{F}-E_{-}\right) -\Theta \left( e_{F}-E_{+}\right) \right]
\label{eq2}
\end{eqnarray}%
and $\left\langle \rho _{+,-}\left( k\right) \right\rangle =\left\langle
\rho _{-,+}\left( k\right) \right\rangle ^{\ast }.$ The following sum rules
follow from the equation of motion%
\begin{equation}
\left\vert \left\langle \rho _{i,i}\left( k\right) \right\rangle \right\vert
^{2}+\left\vert \left\langle \rho _{i,-i}\left( k\right) \right\rangle
\right\vert ^{2}=\left\langle \rho _{i,i}\left( k\right) \right\rangle
\label{sumrule}
\end{equation}%
and 
\begin{equation}
\sum_{j=\pm }\left\langle \rho _{j,j}\left( k\right) \right\rangle
=\sum_{j}\Theta \left( e_{F}-E_{j}\left( k\right) \right) .
\end{equation}

If we write the coherence factor as $\left\langle \rho _{+,-}\left(
k_{1}\right) \right\rangle =\left\vert \left\langle \rho _{+,-}\left(
k_{1}\right) \right\rangle \right\vert e^{i\varphi \left( k_{1}\right) },$
then the internode part of the Fock term in $E_{HF}$ can be written as

\begin{eqnarray}
- &&\frac{1}{4\pi \ell ^{2}L_{z}^{2}}\sum_{k_{1},k_{2}}X\left(
k_{2}-k_{1}\right) \left\vert \left\langle \rho _{+,-}\left( k_{1}\right)
\right\rangle \right\vert  \notag \\
&&\times \left\vert \left\langle \rho _{-,+}\left( k_{2}\right)
\right\rangle \right\vert \cos \left[ \varphi \left( k_{1}\right) -\varphi
\left( k_{2}\right) \right] .
\end{eqnarray}%
The other terms in the Hartree-Fock energy do not depend on the choice of
the phase $\varphi \left( k\right) $ and so it is clear that in the coherent
state, the phase $\varphi \left( k\right) $ must be a constant independent
of $k$ to minimize the energy \textit{i.e.} the internodal excitonic ground
state has a U(1) symmetry.

\section{EXCITONIC STATE WITH $C=1$}

In our numerical calculation for $C=1$, we choose a cutoff $k_{c}\ell =15$
for the wave vector along the $z$ direction so that the dimensionless wave
vector $k\ell \in \left[ -15,15\right] $ in each node. We discretize this
interval into $2N_{p}+1$ points, taking $N_{p}=1000.$ We solve the
self-consistent system of equations (\ref{eq12})-(\ref{eq2}) using an
iterative method. We find that very good convergence is obtained after only $%
100$ iterations if we start the first iteration with the seed $\left\langle
\rho _{\pm ,\pm }\left( k\right) \right\rangle =\Theta \left( \pm
k+k_{F}\right) $ and $\left\langle \rho _{+,-}\left( \pm k_{F}\right)
\right\rangle =\left\langle \rho _{-,+}\left( \pm k_{F}\right) \right\rangle
^{\ast }=1,$ where the Fermi wave vector $k_{F}\ell =4\pi ^{2}\ell ^{3}n_{e}$
is determined by the amount of doping \textit{i.e.} the density of added
electrons $n_{e}.$

To allow coherence in the state $k=0$ in the absence of doping, we remove $%
1/2$ electron at $k=0$ in each node. In order for the excitonic phase to be
the ground state, the cohesive energy $E_{\text{cohe}}=E_{HF}-E_{N}$ must be
negative, where $E_{N}$ is the energy of the normal phase (defined as the
state with Coulomb interaction but without coherence). We study the effect
of three parameters on the excitonic phase: the Fermi velocity $v_{F},$ the
dielectric constant $\varepsilon _{r}$ and the doping level $k_{F}\ell .$

Figure 1 shows the band structure $E_{\pm }\left( k\right) $ in the coherent
(blue lines) and incoherent (black lines) states in the absence of doping.
We have removed a global energy shift $\Sigma _{+}\left( k\right) +\Sigma
_{-}\left( k\right) =-\frac{1}{L_{z}}\sum_{k_{1}}X\left( k_{1}\right) $ (see
Eq. (\ref{sigma})) in both curves to force them to coincide at $k=0.$ The
dashed blue and black lines give the position of the Fermi level for the
corresponding state. For this figure, $v_{F}/c=0.001$ and $\varepsilon
_{r}=1 $ ($c$ is the speed of light in vacuum). All energies are in units of 
$\hslash v_{F}/\ell =7.\,\allowbreak 69\sqrt{B}v_{F}/c$ eV$.$ The
noninteracting band structure (green lines) is modified by the self-energies 
$\Sigma _{\pm }\left( k\right) $ in both the coherent and incoherent states.
The internodal coherence introduces a gap in the band structure. Since there
is only one electron at $k\ell =0$ in the undoped coherent state, the Fermi
level is at the top of the bottom band and the system is insulating. In the
original band structure, the two chiral levels would be separated by the
wave vector $2\mathbf{b}.$ However, since $b$ does not enter our
calculation, we are at liberty to set the origin of both levels at $k=0$ in
all the figures.

\begin{figure}[tbph]
\centering\includegraphics[width = \linewidth]{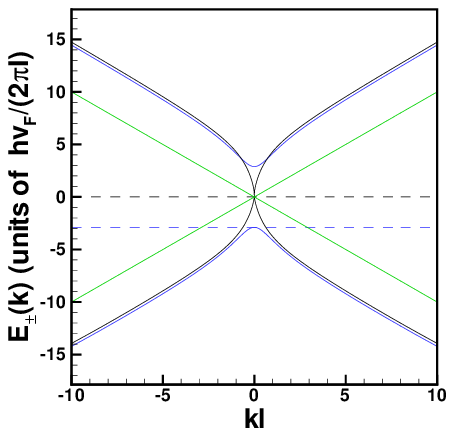}
\caption{Band structure in the undoped coherent and incoherent states. The
blue (black) lines give the electronic dispersion $E_{\pm }\left( k\right) $
in the coherent (incoherent) states while the position of the Fermi level in
each case is given by the dashed line of the corresponding color. A global
energy shift has been removed in both cases to make all the curves centered
at energy $E=0$. The green lines show the noninteracting band structure.
Parameters are $k_{F}\ell =0,v_{F}/c=0.001$ and $\protect\varepsilon _{r}=1.$
}
\label{fig1}
\end{figure}

The corresponding occupations and coherences for the coherent state of Fig.
1 are shown in Fig. 2. They can be compared with the occupation of the $k$
states in the incoherent state which is $\left\langle \rho _{\pm ,\pm
}\left( k\right) \right\rangle =\Theta \left( \mp k\right) =1$ and $%
\left\langle \rho _{+,-}\left( k\right) \right\rangle =0.$ Internodal
coherence leads to a modification of the occupation of the states near $k=0$
where the coherence reaches its maximum value $\left\langle \rho
_{+,-}\left( 0\right) \right\rangle =1/2$ and, as required by the sum rules, 
$\left\langle \rho _{\pm ,\pm }\left( 0\right) \right\rangle =1/2.$ As
expected in a two-level system, the coherence decreases when the difference
in the noninteracting energy $\left\vert e_{+}\left( k\right) -e_{-}\left(
k\right) \right\vert $ increases and so it is maximal at $k=0$ where the two
noninteracting states are degenerate$.$

\begin{figure}[tbph]
\centering\includegraphics[width = \linewidth]{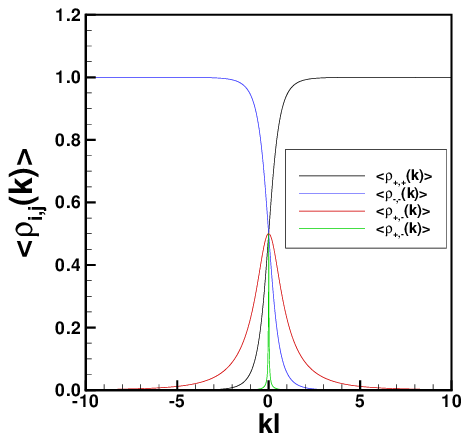}
\caption{Occupations and coherences $\left\langle \protect\rho _{i,j}\left(
k\right) \right\rangle $ in the undoped coherent phase. Parameters are $%
k_{F}\ell =0,v_{F}/c=0.001$ and $\protect\varepsilon _{r}=1$ with the
exception of the green curve where $\protect\varepsilon _{r}=10$ and $%
v_{F}/c=0.002.$ }
\label{fig2}
\end{figure}

We use the integral $\left\langle \rho _{-,+}\right\rangle =\int dk\ell
\left\langle \rho _{-,+}\left( k\ell \right) \right\rangle $ as an order
parameter for the internodal excitonic state. The phase of $\left\langle
\rho _{-,+}\left( k\ell \right) \right\rangle $ being arbitrary, $%
\left\langle \rho _{-,+}\right\rangle $ can be chosen real without any loss
of generality. Figure 3 shows how this quantity depends on the Fermi
velocity $v_{F}/c$ and relative dielectric constant $\varepsilon _{r}.$
Clearly, the coherence decreases rapidly when either one of these parameters
is increased. Indeed, an increase in $\varepsilon _{r}$ decreases the
strength of the Coulomb interaction and an increase in $v_{F}$ increases the
separation in energy of the two levels at $k$ thus decreasing the coherence.
When $v_{F}$ or $\varepsilon _{r}$ increases, $E_{\text{cohe}}\rightarrow 0$
and $\left\langle \rho _{-,+}\right\rangle \rightarrow 0.$

\begin{figure}[tbph]
\centering\includegraphics[width = \linewidth]{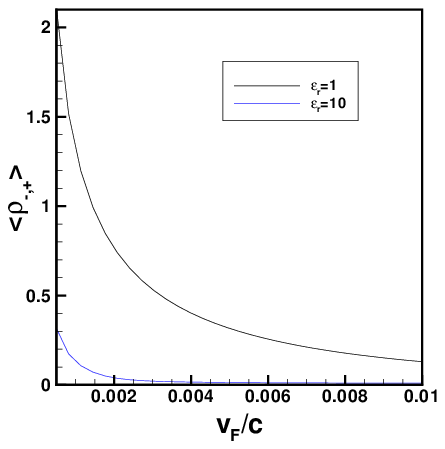}
\caption{Behavior of the order parameter $\left\langle \protect\rho %
_{-,+}\right\rangle $ with the Fermi velocity in the undoped coherent phase
for two different values of the dielectric constants: $\protect\varepsilon %
_{r}=1,10.$ }
\label{fig3}
\end{figure}

We now consider the effect of doping which can be controlled by electric
gating in a WSM. For a single chiral level, the density of states is a
constant given by $g\left( \varepsilon \right) =1/4\pi ^{2}\ell ^{2}\hslash
v_{F}$ so that the Fermi wave vector $k_{F}$ is related to the density of
added electrons per node by $k_{F}\ell =4\pi ^{2}\ell ^{3}n_{e}.$ For $%
k_{F}\ell =0.25,$ the electronic density is $n_{e}=3.\,\allowbreak 75\times
10^{20}B^{\frac{3}{2}}$ e/m$^{3},$ a value that is not atypical in WSMs.

Figure 4 shows the band structure for $k_{F}\ell =0.2$ with $v_{F}/c=0.001$
and $\varepsilon _{r}=1.$ The Fermi level is indicated by the dashed line.
The energy gap at $k=0$ is still present, but the system is now metallic.
The corresponding occupations and coherences are plotted in Fig. 5. No
coherence is possible when a $k$ state is fully occupied in both nodes so
that the occupations are modified only near $\pm k_{F}\ell $ and coherence
occurs only for $\left\vert k\ell \right\vert >k_{F}\ell .$

\begin{figure}[tbph]
\centering\includegraphics[width = \linewidth]{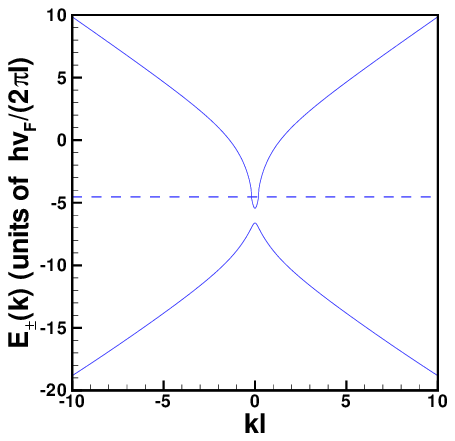}
\caption{Band structure of the doped coherent state. Parameters are $%
k_{F}\ell =0.2,$ $v_{F}/c=0.001$ and $\protect\varepsilon _{r}=1.$ The
position of the Fermi level is indicated by the dashed line. }
\label{fig4}
\end{figure}

\begin{figure}[tbph]
\centering\includegraphics[width = \linewidth]{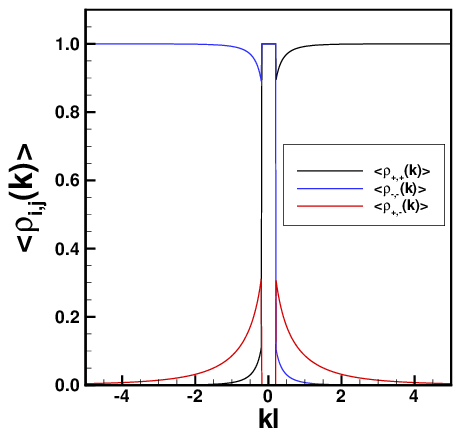}
\caption{Occupations and coherences $\left\langle \protect\rho _{i,j}\left(
k\right) \right\rangle $ in the doped coherent state. Parameters are $%
k_{F}\ell =0.2,v_{F}/c=0.001$ and $\protect\varepsilon _{r}=1.$ }
\label{fig5}
\end{figure}

Figure 6 shows how the order parameter $\left\langle \rho
_{-,+}\right\rangle $ depends on the doping level. An increase in $k_{F}\ell 
$ means that coherence has to be established between two noninteracting
levels with higher energy separation and is consequently weaker. By
electron-hole symmetry of the original band structure, the same results are
obtained for hole doping.

\begin{figure}[tbph]
\centering\includegraphics[width = \linewidth]{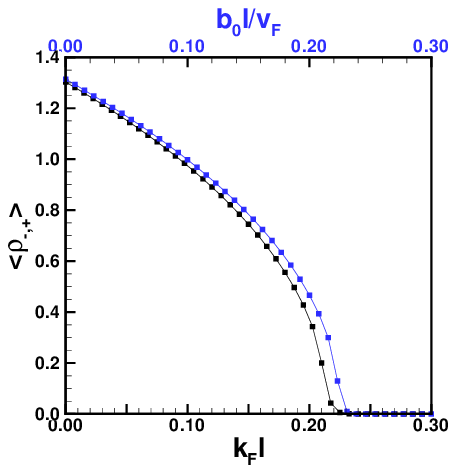}
\caption{Behavior of the order parameter $\left\langle \protect\rho %
_{-,+}\right\rangle $ with electronic doping $k_{F}\ell $ in the coherent
phase (bottom axis) and with the biais $b_{0}$ (top axis). Parameters are $%
v_{F}/c=0.001$ and $\protect\varepsilon _{r}=1.$ }
\label{fig6}
\end{figure}

The results of this section show that the excitonic state can be realized
with realistic values of the Fermi velocity, dielectric constant and doping.
However, it is fragile and disappears rapidly as these parameters are
increased.

At this point, we must say a word about the validity of our approximations.
In order to restrict the Hilbert space to the chiral levels, we need the
Coulomb interaction to be small with respect to the energy gap between the $%
n=0$ and $n=\pm 1$ Landau levels of the noninteracting electron gas. That
is, we must ensure that $e^{2}/4\pi \varepsilon _{0}\varepsilon _{r}\ell <%
\sqrt{2}\hslash v_{F}/\ell $ \textit{i.e.} $\alpha /\varepsilon _{r}%
\overline{v}_{F}\ll \sqrt{2}$ where $\alpha =e^{2}/4\pi \varepsilon
_{0}\hslash c$ is the fine-structure constant and $\overline{v}_{F}=v_{F}/c.$
Since $\varepsilon _{r}$ can be large in a WSM, this condition can be
satisfied in principle. However, as our calculation shows, the coherence $%
\left\langle \rho _{-,+}\right\rangle $ decreases rapidly with the product $%
\varepsilon _{r}\overline{v}_{F}.$ Nevertheless, Fig. 3 shows that there is
a range of values of $\varepsilon _{r}\overline{v}_{F}$ where this condition
is satisfied and coherence is possible.

The calculation of the self-energy implies integration over the wave vector $%
k$ and so we must ensure that the cutoff wave vector $k_{c}$ is such that
the energy in the chiral band is lower than that of the $n=1$ Landau level
which implies $k_{c}\ell <\sqrt{2}.$ This requires that the modification of
the occupations from their noninteracting value be negligible for $k\ell >%
\sqrt{2}.$ As Fig. 2 shows, this is not the case for $\varepsilon _{r}=1$
and $\overline{v}_{F}=0.001$ but the condition is satisfied for $\varepsilon
_{r}=10$ and $\overline{v}_{F}=0.002$ (the green curve).

When doping is considered, the Fermi level must also be below the $n=1$
Landau band. This condition is satisfied in our numerical calculation since,
as shown in Fig. 6, the coherence $\left\langle \rho _{-,+}\right\rangle $
decreases to zero well before $k_{F}\ell =\sqrt{2}$ is reached.

Our calculation assumes $T=0$ K. In the absence of doping and for the set of
parameters $v_{F}/c=0.001,\varepsilon _{r}=1$\ and $B=10$T, the difference
in the energy per electron between the coherent and incoherent states (the
cohesive energy) is approximately $4$ K while for $v_{F}/c=0.002,\varepsilon
_{r}=10$\ and $B=10$ T, where the coherence is much smaller (see Fig. 3),
this difference decreases to $4$ mK. If we use these figures to approximate
the melting temperature for the coherent state, then we can conclude that
the coherence should survive at finite, but small, temperature.

Although the discovery of a WSM\ with a single pair of Weyl nodes, Eu$_{3}$In%
$_{2}$As$_{4},$\ has been reported recently\cite{Jia2024} (and others should
exist according to $ab$\ $initio$\ calculations), most WSMs\ have more than
two Weyl points. Because the internodal coherence occurs in momentum space
and not in real space, we believe that it should remain possible in WSMs
with more pairs of nodes. That is, if their Dirac points are at the same
energy. If they are not, then the doping of the different nodes will be
different and the coherence will probably be lost. To verify this, we
calculated the effect of adding an electrical bias that shifts the energy of
the two nodes. In our model, this is done by adding the term $-\tau \hslash
b_{0}$ to the single-particle energy in Eq. (2) and the term $\hslash b_{0}$
to $e(k)$ in Eq. (24). In the incoherent state, $\left\langle \rho
_{+,+}\left( k\right) \right\rangle =1$ for $k\ell \in \left[ -2\frac{%
b_{0}\ell }{v_{F}},k_{c}\ell \right] $ and $\left\langle \rho _{-,-}\left(
k\right) \right\rangle =1$ for $k\ell \in \left[ -k_{c}\ell ,0\right] $ so
that there is a difference in density of added electrons given by $\Delta
n_{e}=b_{0}/2\pi ^{2}v_{F}\ell ^{2}$ between the two nodes. Coherence cannot
be established in the region $k\ell \in \left[ -2\frac{b_{0}\ell }{v_{F}},0%
\right] $ where the two chiral levels are occupied. It can only occur in the
flanks of this region which is indeed what we find numerically. The blue
line in Fig. 6 shows how the order parameter decreases with $b_{0}.$ As
expected the coherence decreases with the electrical bias. For $B=10$ T and $%
v_{F}/c=0.001,$ it vanishes for $\hslash b_{0}\gtrsim 6.3$ meV. Note that we
assume that the two nodes are at equilibrium so that they share the same
Fermi level.

In concluding this section, we remark that the excitonic state in our
description is uniform spatially since we consider the two nodes as separate
systems and write the total density as $n\left( \mathbf{r},z\right)
=\sum_{\tau }\Psi _{\tau }^{\dag }\left( \mathbf{r},z\right) \Psi _{\tau
}\left( \mathbf{r},z\right) .$ But, when the two nodes are considered as one
system, the density should be written as $n\left( \mathbf{r},z\right) =\Psi
^{\dag }\left( \mathbf{r},z\right) \Psi \left( \mathbf{r},z\right) $ with
the field operator defined by

\begin{eqnarray}
\Psi \left( \mathbf{r},z\right) &=&w_{0,k,X}\left( \mathbf{r}_{\bot
},z\right) e^{-ibz}c_{k,X,-}  \label{grandpsi} \\
&&+w_{0,k,X}\left( \mathbf{r}_{\bot },z\right) e^{ibz}c_{k,X,+},  \notag
\end{eqnarray}%
where the summation over $k$ is restricted to the small momentum region near
each node where the dispersion is linear. Performing an integration over $%
\mathbf{r}_{\bot },$ we have for the average density along the $z$ direction
apart from an unimportant constant%
\begin{equation}
\left\langle n\left( z\right) \right\rangle \sim \frac{S}{2\pi ^{2}\ell ^{3}}%
\cos \left( 2bz+\varphi \right) \left\langle \rho _{-,+}\right\rangle ,
\end{equation}%
where $\varphi $ is the U(1) phase of the complex order parameter $%
\left\langle \rho _{-,+}\right\rangle .$ In this description, the excitonic
phase is also a charge density wave state which is modulated by the axion
wave vector $b$ and whose amplitude depends on the magnetic field and order
parameter.

\section{EXCITONIC STATES FOR CHERN NUMBER $C=2$ AND$\ C=3$}

For Weyl nodes with Chern number $C=2,3,$ there are respectively $4$ and $6$
quantum states to consider. We denote them by the super-indices $I,J=\left(
n,\tau \right) =1,2,...,6$ with the correspondence 
\begin{eqnarray}
1 &=&\left( 0,+\right) ;2=\left( 0,-\right) ,  \notag \\
3 &=&\left( 1,+\right) ;4=\left( 1,-\right) , \\
5 &=&\left( 2,+\right) ;6=\left( 2,-\right) .  \notag
\end{eqnarray}%
We solve the equation of motion for the Green's function given in Eq. (\ref%
{eqgreen}). The components of the $F$ matrix are defined by 
\begin{equation}
F_{I,J}\left( k\right) =\frac{1}{\hslash }\left[ \left( -\tau \hslash
v_{F}k-\mu \right) \delta _{I,J}-\Sigma _{I,J}\left( k\right) \right] ,
\end{equation}%
where the self-energies for $C=1,2$ are given by%
\begin{eqnarray}
\Sigma _{n,n}^{\left( \tau ,\tau ^{\prime }\right) }\left( k\right) &=&-%
\frac{1}{L_{z}}\sum_{k_{1}}\sum_{n_{1}}X_{n_{1},n,n,n_{1}}\left(
k-k_{1}\right)  \notag \\
&&\times \left\langle \rho _{n_{1},n_{1}}^{\left( \tau ,\tau ^{\prime
}\right) }\left( k_{1}\right) \right\rangle
\end{eqnarray}%
and%
\begin{eqnarray}
\Sigma _{n,n^{\prime }\neq n}^{\left( \tau ,\tau ^{\prime }\right) }\left(
k\right) &=&-\frac{1}{L_{z}}\sum_{k_{1}}X_{n^{\prime },n^{\prime
},n,n}\left( k-k_{1}\right)  \notag \\
&&\times \left\langle \rho _{n^{\prime },n}^{\left( \tau ^{\prime },\tau
\right) }\left( k_{1}\right) \right\rangle .
\end{eqnarray}%
For $C=3,$ however, there are four additional contributions to some of the
self-energies $\Sigma _{n,n^{\prime }\neq n}^{\left( \tau ,\tau ^{\prime
}\right) }\left( k\right) $ which are given by%
\begin{eqnarray}
\Sigma _{0,1}^{\left( \tau ,\tau ^{\prime }\right) }\left( k\right)
&\rightarrow &-\frac{1}{L_{z}}\sum_{k_{1}}X_{2,1,0,1}\left( k-k_{1}\right) 
\notag \\
&&\times \left\langle \rho _{2,1}^{\left( \tau ^{\prime },\tau \right)
}\left( k_{1}\right) \right\rangle , \\
\Sigma _{1,0}^{\left( \tau ,\tau ^{\prime }\right) }\left( k\right)
&\rightarrow &-\frac{1}{L_{z}}\sum_{k_{1}}X_{1,0,1,2}\left( k-k_{1}\right) 
\notag \\
&&\times \left\langle \rho _{1,2}^{\left( \tau ^{\prime },\tau \right)
}\left( k_{1}\right) \right\rangle ,
\end{eqnarray}%
and 
\begin{eqnarray}
\Sigma _{1,2}^{\left( \tau ,\tau ^{\prime }\right) }\left( k\right)
&\rightarrow &-\frac{1}{L_{z}}\sum_{k_{1}}X_{1,2,1,0}\left( k-k_{1}\right) 
\notag \\
&&\times \left\langle \rho _{1,0}^{\left( \tau ^{\prime },\tau \right)
}\left( k_{1}\right) \right\rangle , \\
\Sigma _{2,1}^{\left( \tau ,\tau ^{\prime }\right) }\left( k\right)
&\rightarrow &-\frac{1}{L_{z}}\sum_{k_{1}}X_{0,1,2,1}\left( k-k_{1}\right) 
\notag \\
&&\left\langle \rho _{0,1}^{\left( \tau ^{\prime },\tau \right) }\left(
k_{1}\right) \right\rangle .
\end{eqnarray}

Figure 7(a) shows the occupations and coherences and Fig. 7(b) the
corresponding band structure for $C=2$ in the excitonic state for $%
v_{F}/c=0.001$ and $\varepsilon _{r}=1$. The band structure in the absence
of coherence but with interaction is shown in the inset of Fig. 7(b) for $%
C=2.$ The $n=0,1$ bands have different self-energies. They are thus shifted
differently in energy creating many degeneracy points. As with $C=1,$ the
coherences gap the whole band structure. For Fermi velocity $v_{F}/c\in %
\left[ 0.001,0.1\right] ,$ only internodal coherence in the same band is
present. It decreases with $v_{F}/c$ as shown in the inset of Fig. 7(d) and,
at $v_{F}/c\approx 0.011,$ it drops abruptly to zero where it is replaced by
entanglement between states $\left( 1,4\right) $ and $\left( 2,3\right) $ as
shown in Fig. 7(c). The corresponding band structure after this phase
transition is shown in Fig. 7(d). It is modified from the interacting but
incoherent band structure shown in the inset of Fig. 7(b) but only in a very
small range of wave vector $k\ell$.

\begin{figure*}[tbph]
\centering\includegraphics[width = \linewidth]{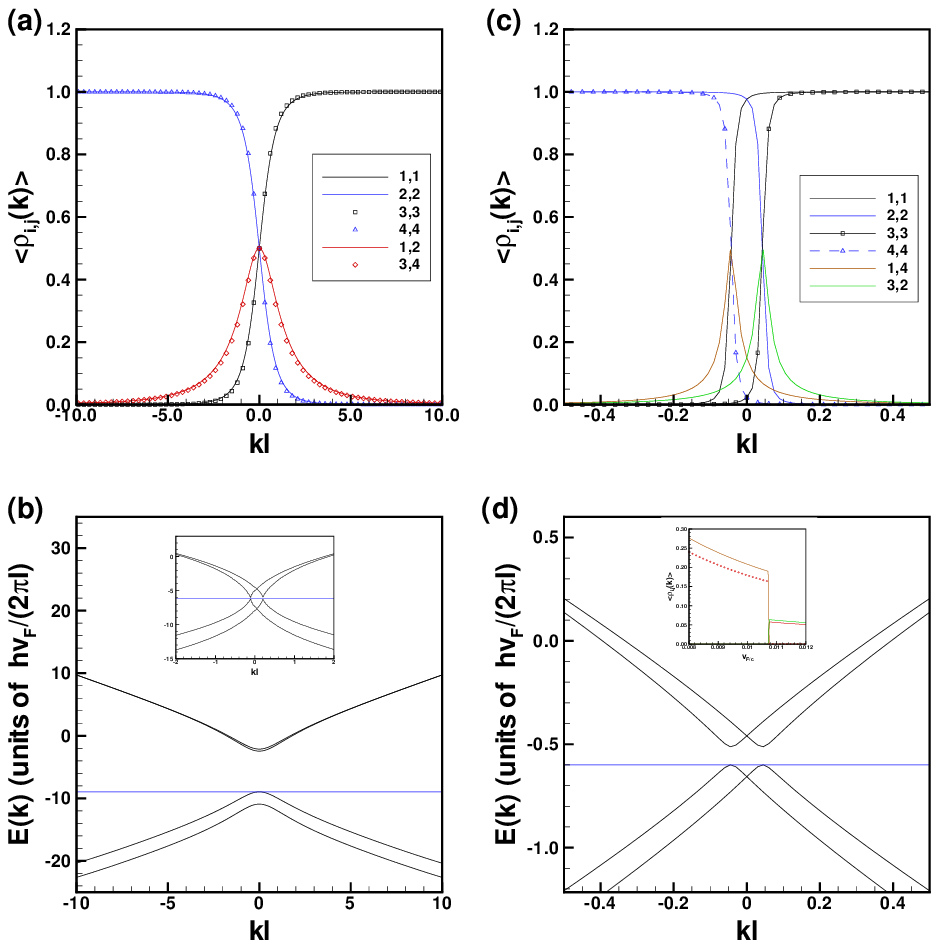}
\caption{Excitonic state for $C=2.$ (a) occupations and coherences and (b)
band structure for $v_{F}/c=0.001.$ Pannels (c) and (d) show the same but
for $v_{F}/c=0.01$. The inset in (b) shows the gapless band structure in the
absence of coherence and that in (d) shows the behavior of the coherences
with $v_{F}/c$ (the lines legend is as in (c)). The blue lines in (b) and
(d) indicate the position of the Fermi level. Parameters are $k_{F}\ell =0,$
and $\protect\varepsilon _{r}=1.$}
\label{fig7}
\end{figure*}

Figure 8(a) shows the occupations and coherences and Fig. 8(b) the
corresponding band structure for $C=3$ in the excitonic state for $%
v_{F}/c=0.001$ and $\varepsilon _{r}=1$. As for $C=2,$ only internodal
coherence in the same band is present at this Fermi velocity and the band
structure is gapped. We find that this ground state persists up to $%
v_{F}/c\approx 0.006$ where there is a transition to a different type of
coherent state that we were not able to identify completely, the number of
such states being quite large for $C=3.$

\begin{figure}[tbph]
\centering\includegraphics[width = \linewidth]{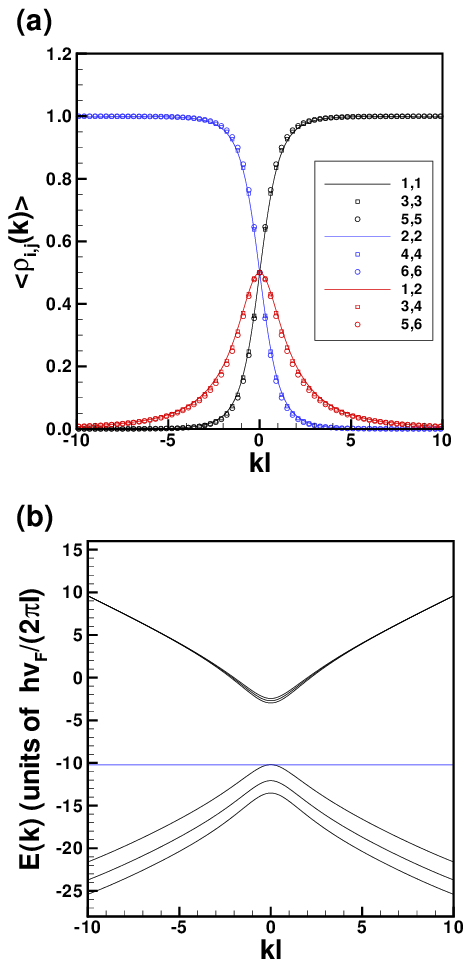}
\caption{Excitonic state for $C=3.$ (a) occupations and coherences and (b)
band structure. The blue line in (b) indicates the position of the Fermi
level. Parameters are $v_{F}/c=0.001,k_{F}\ell =0$ and $\protect\varepsilon %
_{r}=1.$ }
\label{fig8}
\end{figure}

\begin{widetext}

When only internodal coherence in the same level is present, the ground
state energy per volume $V$ is given by

\begin{eqnarray}
\frac{E_{HF}}{V} &=&\frac{1}{2\pi \ell ^{2}}\frac{1}{L_{z}}\sum_{n,k,\tau
}e_{\tau }\left( k\right) \left\langle \rho _{n,n}^{\left( \tau ,\tau
\right) }\left( k\right) \right\rangle  \\
&&-\frac{1}{2\pi \ell ^{2}}\frac{1}{L_{z}^{2}}\sum_{\tau
}\sum_{k_{1},k_{2}}\sum_{n_{1},n_{2}}X_{n_{1},n_{2},n_{2},n_{1}}\left(
k_{2}-k_{1}\right) \left\langle \rho _{n_{1},n_{1}}^{\left( \tau ,\tau
\right) }\left( k_{1}\right) \right\rangle \left\langle \rho
_{n_{2},n_{2}}^{\left( \tau ,\tau \right) }\left( k_{2}\right) \right\rangle 
\notag \\
&&-\frac{2}{2\pi \ell ^{2}}\frac{1}{L_{z}^{2}}\sum_{k_{1},k_{2}}%
\sum_{n_{1}}X_{n_{1},n_{1},n_{1},n_{1}}\left( k_{2}-k_{1}\right) \left\vert
\left\langle \rho _{n_{1},n_{1}}^{\left( +,-\right) }\left( k_{1}\right)
\right\rangle \right\vert \left\vert \left\langle \rho
_{n_{1},n_{1}}^{\left( -,+\right) }\left( k_{2}\right) \right\rangle
\right\vert \cos \left( \varphi _{n_{1}}\left( k_{1}\right) -\varphi
_{n_{1}}\left( k_{2}\right) \right)   \notag \\
&&-\frac{2}{2\pi \ell ^{2}}\frac{1}{L_{z}^{2}}\sum_{k_{1},k_{2}}%
\sum_{n_{1},n_{2}\neq n_{1}}X_{n_{1},n_{2},n_{2},n_{1}}\left(
k_{2}-k_{1}\right) \left\vert \left\langle \rho _{n_{1},n_{1}}^{\left(
+,-\right) }\left( k_{1}\right) \right\rangle \right\vert \left\vert
\left\langle \rho _{n_{2},n_{2}}^{\left( -,+\right) }\left( k_{2}\right)
\right\rangle \right\vert \cos \left( \varphi _{n_{1}}\left( k_{1}\right)
-\varphi _{n_{2}}\left( k_{2}\right) \right) ,  \notag
\end{eqnarray}%
with the phases defined by $\left\langle \rho _{n,n}^{\left( +,-\right)
}\left( k\right) \right\rangle =\left\vert \left\langle \rho _{n,n}^{\left(
+,-\right) }\left( k\right) \right\rangle \right\vert e^{i\varphi _{n}\left(
k\right) }.$ The energy is minimized when $\varphi _{n_{1}}\left(
k_{1}\right) =\varphi _{n_{1}}\left( k_{2}\right) $ and $\varphi
_{n_{1}}\left( k_{1}\right) =\varphi _{n_{2}}\left( k_{2}\right) .$ Thus,
all internodal coherent states for $C=1,2,3$ are invariant with respect to
one global phase.

\end{widetext}

\section{GENERAL EQUATION FOR THE RESPONSE FUNCTIONS}

In order to derive the response functions in the excitonic state we compute
the two-particle Matsubara Green's functions 
\begin{eqnarray}
L_{a,b,c,d}\left( 1,2,3,4\right) &=&-\left\langle T\Psi _{a}^{\dagger
}\left( 1\right) \Psi _{b}\left( 2\right) \Psi _{c}^{\dagger }\left(
3\right) \Psi _{d}\left( 4\right) \right\rangle  \notag \\
&&+G_{b,a}\left( 2,1\right) G_{d,c}\left( 4,3\right) ,
\end{eqnarray}%
where the single-particle Matsubara Green's function is defined by%
\begin{equation}
G_{a,b}\left( 1,2\right) =-\left\langle T\Psi _{a}\left( 1\right) \Psi
_{b}^{\dag }\left( 2\right) \right\rangle .
\end{equation}%
The numbers refer to the position vector and imaginary time\textit{\ i.e.} $%
1=\left( \mathbf{u}_{1},\tau _{1}\right) ,$ the integral $\int d\overline{1}%
=\int_{0}^{\beta \hslash }d\tau _{1}\int d^{3}r_{1}$ and $a,b,c,d$ are node
indices.

The single-particle Green's function introduced in the previous section was
computed in the Hartree-Fock approximation which is defined by%
\begin{gather}
G_{a,b}\left( 1,2\right) =G_{a,b}^{0}\left( 1,2\right)  \notag \\
+\sum_{c,d}\int d\overline{3}G_{a,c}^{0}\left( 1,\overline{3}\right) \Sigma
_{c,d}^{HF}\left( \overline{3},\overline{4}\right) G_{d,b}\left( \overline{4}%
,2\right) ,
\end{gather}%
where $G_{a,c}^{0}$ is a non-interacting Green's function and the
Hartree-Fock self-energy is defined by%
\begin{eqnarray}
\Sigma _{c,d}\left( 5,6\right) &=&\frac{1}{\hslash }\delta _{c,d}\int d%
\overline{7}\delta \left( 5-6\right) V\left( 5-\overline{7}\right)
G_{g,g}\left( \overline{7},\overline{7}^{+}\right)  \notag \\
&&-\frac{1}{\hslash }V\left( 5-6\right) G_{c,d}\left( 5,6\right) ,
\label{selff}
\end{eqnarray}%
where $V\left( 1-2\right) =V\left( \mathbf{u}_{1}-\mathbf{u}_{2}\right)
\delta \left( \tau _{1}-\tau _{2}\right) $ is the Coulomb interaction which
is independent of the node index. The two terms on the right-hand side of
Eq. (\ref{selff}) are respectively the Hartree and Fock self-energies.

We derive the two-particle Green's function in the generalized random-phase
approximation (GRPA) which consists in the summation of bubble and ladder
diagrams. The GRPA is obtained by a functional derivative of the
single-particle Green's function and is a conserving approximation\cite{Baym}%
. More precisely, it is defined by the equation

\begin{widetext} 

\begin{gather}
L_{a,b,c,d}\left( 1,2,3,4\right) =G_{b,c}\left( 2,3\right) G_{d,a}\left(
4,1\right)   \notag \\
+\sum_{e,g}\frac{1}{\hslash }\int d\overline{5}\int d\overline{6}%
G_{b,e}\left( 2,\overline{5}\right) G_{e,a}\left( \overline{5},1\right)
V\left( \overline{5}-\overline{6}\right) L_{g,g,c,d}\left( \overline{6^{+}},%
\overline{6},3,4\right)   \notag \\
-\sum_{e,f}\frac{1}{\hslash }\int d\overline{5}\int d\overline{6}%
G_{b,e}\left( 2,\overline{5}\right) G_{f,a}\left( \overline{6},1\right)
V\left( \overline{5}-\overline{6}\right) L_{f,e,c,d}\left( \overline{6},%
\overline{5},3,4\right) .  \label{llgrpa}
\end{gather}

\end{widetext} 

This equation couples all 16 Green's functions together and we can extract
from it the two-particle Green's functions 
\begin{eqnarray}
&&P_{k_{1},k_{2},k_{3},k_{4}}^{\tau _{a},\tau _{b},\tau _{c},\tau
_{d}}\left( \mathbf{q}_{\bot },\mathbf{q}_{\bot }^{\prime };\tau \right) \\
&=&-N_{\varphi }\left\langle T\rho _{\tau _{a},\tau _{b}}\left( \mathbf{q}%
_{\bot },k_{1},k_{2};\tau \right) \rho _{\tau _{c},\tau _{d}}\left( -\mathbf{%
q}_{\bot }^{\prime },k_{3},k_{4};0\right) \right\rangle ,  \notag
\end{eqnarray}%
where the operators $\rho _{\tau _{a},\tau _{b}}\left( \mathbf{q}_{\bot }%
\mathbf{,}k,k^{\prime };\tau \right) $ now depends on imaginary time and are
defined by%
\begin{eqnarray}
\rho _{\tau _{a},\tau _{b}}\left( \mathbf{q}_{\bot }\mathbf{,}k,k^{\prime
};\tau \right) &=&\frac{1}{N_{\varphi }}\sum_{X}e^{-iq_{x}X}e^{iq_{x}q_{y}%
\ell ^{2}/2}c_{k,X,\tau _{a}}^{\dagger }\left( \tau \right)  \notag \\
&&\times c_{k^{\prime },X-q_{y}\ell ^{2},\tau _{b}}\left( \tau \right) .
\end{eqnarray}

We calculate the two-particle Green's function in the uniform state so that
only the occupation and coherences $\left\langle \rho _{\tau _{a},\tau
_{b}}\left( \mathbf{q}_{\bot }=0\mathbf{,}k,k\right) \right\rangle
=\left\langle \rho ^{\tau _{a},\tau _{b}}\left( k\right) \right\rangle $ are
nonzero. Moreover, we restrict our analysis to response functions with wave
vectors along the direction of the magnetic field \textit{i.e.} take $%
\mathbf{q}_{\bot }=\mathbf{q}_{\bot }^{\prime }=0$ in $%
P_{k_{1},k_{2},k_{3},k_{4}}^{\tau _{a},\tau _{b},\tau _{c},\tau _{d}}\left( 
\mathbf{q}_{\bot },\mathbf{q}_{\bot }^{\prime };\tau \right) .$

One of us (R. C.) has given in appendix A of Ref. \onlinecite{Bertrand2019}
a detailed derivation of the GRPA equation of motion for $L_{a,b,c,d}\left(
1,2,3,4\right) $ and $P_{k_{1},k_{2},k_{3},k_{4}}^{\tau _{a},\tau _{b},\tau
_{c},\tau _{d}}\left( 0,0;i\Omega _{n}\right) ,$ where $\Omega _{n}=2n\pi
/\beta \hslash $ with $n=0,\pm 1,\pm 2,...$is a bosonic Matsubara frequency.
We refer the reader to this appendix and give here only the main results.

Hereafter, we restrict our analysis to nodes with Chern number $C=1$ in
order for the size of the matrices that are involved in the calculation to
be manageable numerically. We first define a matrix containing the 16
response functions%
\begin{equation}
P_{k+Q,k}^{I,J}\left( \omega \right) =\sum_{k^{\prime }}P_{k+Q,k,k^{\prime
},k^{\prime }+Q}^{I,J}\left( \omega \right) ,
\end{equation}%
where the super-indices $I,J=1,2,3,4$ are defined in the following way: for
the rows $I=\left( \tau _{a},\tau _{b}\right) =\left( +,+\right) ,\left(
+,-\right) ,\left( -,+\right) ,\left( -,-\right) $ and, for the columns $%
J=\left( \tau _{c},\tau _{d}\right) =\left( +,+\right) ,\left( -,+\right)
,\left( +,-\right) ,\left( -,-\right) .$ We also define the matrices

\begin{widetext} 
\begin{equation}
\overline{E}\left( k,Q\right) =\frac{1}{\hslash }\left( 
\begin{array}{cccc}
\left( e_{+}\left( k\right) -e_{+}\left( k+Q\right) \right)  & 0 & 0 & 0 \\ 
0 & e_{-}\left( k\right) -e_{+}\left( k+Q\right)  & 0 & 0 \\ 
0 & 0 & e_{+}\left( k\right) -e_{-}\left( k+Q\right)  & 0 \\ 
0 & 0 & 0 & e_{-}\left( k\right) -e_{-}\left( k+Q\right) 
\end{array}%
\right) 
\end{equation}%
and 
\begin{equation}
\overline{\Sigma }\left( k,Q\right) =\frac{1}{\hslash }\left( 
\begin{array}{cccc}
\left( \Sigma _{+}\left( k\right) -\Sigma _{+}\left( k+Q\right) \right)  & 
\Sigma \left( k\right)  & -\Sigma \left( k+Q\right)  & 0 \\ 
\Sigma \left( k\right)  & \Sigma _{-}\left( k\right) -\Sigma \left(
k+Q\right)  & 0 & -\Sigma \left( k+Q\right)  \\ 
-\Sigma \left( k+Q\right)  & 0 & \Sigma _{+}\left( k\right) -\Sigma
_{-}\left( k+Q\right)  & \Sigma \left( k\right)  \\ 
0 & -\Sigma \left( k+Q\right)  & \Sigma \left( k\right)  & \Sigma _{-}\left(
k\right) -\Sigma _{-}\left( k+Q\right) 
\end{array}%
\right) 
\end{equation}%
and 
\begin{equation}
\overline{B}\left( k,Q\right) =\left( 
\begin{array}{cccc}
\begin{array}{c}
\left\langle \rho _{+,+}\left( k+Q\right) \right\rangle  \\ 
-\left\langle \rho _{+,+}\left( k\right) \right\rangle 
\end{array}
& -\left\langle \rho _{-,+}\left( k\right) \right\rangle  & \left\langle
\rho _{+,-}\left( k+Q\right) \right\rangle  & 0 \\ 
-\left\langle \rho _{+,-}\left( k\right) \right\rangle  & 
\begin{array}{c}
\left\langle \rho _{+,+}\left( k+Q\right) \right\rangle  \\ 
-\left\langle \rho _{-,-}\left( k\right) \right\rangle 
\end{array}
& 0 & \left\langle \rho _{+,-}\left( k+Q\right) \right\rangle  \\ 
\left\langle \rho _{-,+}\left( k+Q\right) \right\rangle  & 0 & 
\begin{array}{c}
\left\langle \rho _{-,-}\left( k+Q\right) \right\rangle  \\ 
-\left\langle \rho _{+,+}\left( k\right) \right\rangle 
\end{array}
& -\left\langle \rho _{-,+}\left( k\right) \right\rangle  \\ 
0 & \left\langle \rho _{-,+}\left( k+Q\right) \right\rangle  & -\left\langle
\rho _{+,-}\left( k\right) \right\rangle  & 
\begin{array}{c}
\left\langle \rho _{-,-}\left( k+Q\right) \right\rangle  \\ 
-\left\langle \rho _{-,-}\left( k\right) \right\rangle 
\end{array}%
\end{array}%
\right) 
\end{equation}%
\end{widetext} 

and these other matrices%
\begin{equation}
\overline{H}\left( Q\right) =\frac{e^{2}}{\varepsilon _{0}\hslash }\left( 
\begin{array}{cccc}
H\left( Q\right) & 0 & 0 & H\left( Q\right) \\ 
0 & 0 & 0 & 0 \\ 
0 & 0 & 0 & 0 \\ 
H\left( Q\right) & 0 & 0 & H\left( Q\right)%
\end{array}%
\right)
\end{equation}%
and%
\begin{equation}
\overline{X}\left( k\right) =\frac{e^{2}}{\varepsilon _{0}\hslash }\left( 
\begin{array}{cccc}
X\left( k\right) & 0 & 0 & 0 \\ 
0 & X\left( k\right) & 0 & 0 \\ 
0 & 0 & X\left( k\right) & 0 \\ 
0 & 0 & 0 & X\left( k\right)%
\end{array}%
\right) .
\end{equation}%
where 
\begin{eqnarray}
H\left( Q\right) &=&\frac{1}{2\pi }\frac{1}{\left( Q\ell \right) ^{2}}, \\
X\left( k\right) &=&\frac{1}{4\pi }e^{\frac{\left( k\ell \right) ^{2}}{2}%
}\Gamma \left( 0,\frac{\left( k\ell \right) ^{2}}{2}\right) .
\end{eqnarray}

We can then write the GRPA system of equations for the response function in
the compact matrix form

\begin{eqnarray}
&&\left[ I\left( i\Omega _{n}\right) -\overline{E}\left( k,Q\right) -%
\overline{\Sigma }\left( k,Q\right) \right] \overline{P}_{k+Q,k}\left(
i\Omega _{n}\right)  \notag \\
&=&\overline{B}\left( k,Q\right)  \notag \\
&&+\overline{B}\left( k,Q\right) \overline{H}\left( Q\right) \frac{1}{L_{z}}%
\sum_{k^{\prime }}\overline{P}_{k^{\prime }+Q,k^{\prime }}\left( i\Omega
_{n}\right)  \notag \\
&&-\frac{1}{L_{z}}\sum_{k^{\prime }}\overline{B}\left( k,Q\right) \overline{X%
}\left( k-k^{\prime }\right) \overline{P}_{k^{\prime }+Q,k^{\prime }}\left(
i\Omega _{n}\right) .  \label{grpa3}
\end{eqnarray}%
Once all the elements of the matrix $\overline{P}_{k+Q,k}\left( \omega
\right) $ have been calculated, we obtain the retarded responses by making
the analytic continuation $i\Omega _{n}\rightarrow \omega +i\delta $ and
then summing over $k$ \textit{i.e.} 
\begin{equation}
\overline{P}\left( \omega ,Q\right) =\frac{N_{\varphi }}{SL_{z}}\sum_{k}%
\overline{P}_{k+Q,k}\left( \omega \right) .
\end{equation}

We also use below the proper response functions, $\overline{\widetilde{P}}%
_{k+Q,k}\left( i\Omega _{n}\right) ,$ which are defined by the summation of
the connected diagrams only \textit{i.e.} by setting $\overline{H}\left(
Q\right) $ in Eq. (\ref{grpa3}) so that%
\begin{eqnarray}
&&\frac{1}{L_{z}}\sum_{k^{\prime }}\left[ \left( I\left( i\Omega _{n}\right)
-\overline{E}\left( k,Q\right) -\overline{\Sigma }\left( k,Q\right) \right)
L_{z}\delta _{k,k^{\prime }}\right.  \notag \\
&&\left. +\overline{B}\left( k,Q\right) \overline{X}\left( k-k^{\prime
}\right) \right] \overline{\widetilde{P}}_{k^{\prime }+Q,k^{\prime }}\left(
i\Omega _{n}\right)  \notag \\
&=&\overline{B}\left( k,Q\right) .
\end{eqnarray}%
In terms of the proper response functions, the GRPA equation can be written
as%
\begin{eqnarray}
\overline{P}_{k+Q,k}\left( i\Omega _{n}\right) &=&\overline{\widetilde{P}}%
_{k+Q,k}\left( i\Omega _{n}\right) +\overline{\widetilde{P}}_{k+Q,k}\left(
i\Omega _{n}\right)  \notag \\
&&\times \overline{H}\left( Q\right) \frac{1}{L_{z}}\sum_{k^{\prime }}%
\overline{P}_{k^{\prime }+Q,k^{\prime }}\left( \omega \right) .
\end{eqnarray}

\section{DENSITY, CURRENT AND\ EXCITONIC RESPONSE FUNCTIONS}

In this section, we define the density, current and excitonic response
functions. The Fourier transform $n\left( \mathbf{q}_{\bot },q_{z}\right) $
of the second quantized charge density operator is given by%
\begin{eqnarray}
n\left( 0,Q\right) &=&-e\sum_{\tau }\int d\mathbf{u}\Psi _{\tau }^{\dag
}\left( \mathbf{u}\right) e^{-iQz}\Psi _{\tau }\left( \mathbf{u}\right) 
\notag \\
&=&-eN_{\varphi }\sum_{\tau ,k}\rho _{\tau ,\tau }\left( \mathbf{q}_{\bot
}=0,k,k+Q\right) ,
\end{eqnarray}%
while, with the current operator given by $j_{z,\tau }=-e\tau v_{F}\sigma
_{z},$ the second-quantized form $J_{z,\tau }\left( \mathbf{q}_{\bot
},q_{z}\right) $ is%
\begin{eqnarray}
J_{z,\tau }\left( 0,Q\right) &=&\int d\mathbf{u}\Psi _{\tau }^{\dag }\left( 
\mathbf{u}\right) e^{-iQz}j_{z,\tau }\Psi _{\tau }\left( \mathbf{u}\right) 
\notag \\
&=&ev_{F}N_{\varphi }\sum_{\tau ,k}\tau \rho _{\tau ,\tau }\left( \mathbf{q}%
_{\bot }=0,k,k+Q\right) .
\end{eqnarray}%
Thus, the density, $\chi _{nn},$ and current, $\chi _{jj},$ response
functions are given in the GRPA by

\begin{eqnarray}
\chi _{nn}\left( \omega ,Q\right) &=&e^{2}\left[ P^{1,1}\left( \omega
,Q\right) +P^{1,4}\left( \omega ,Q\right) \right]  \notag \\
&&+e^{2}\left[ P^{4,1}\left( \omega ,Q\right) +P^{4,4}\left( \omega
,Q\right) \right]
\end{eqnarray}%
and 
\begin{eqnarray}
\chi _{jj}\left( \omega ,Q\right) &=&e^{2}v_{F}^{2}\left[ P^{1,1}\left(
\omega ,Q\right) -P^{1,4}\left( \omega ,Q\right) \right]  \notag \\
&&+e^{2}v_{F}^{2}\left[ -P^{4,1}\left( \omega ,Q\right) +P^{4,4}\left(
\omega ,Q\right) \right]
\end{eqnarray}%
respectively. We define an excitonic response function by%
\begin{equation}
\chi _{\text{exc}}\left( \omega ,Q\right) =P^{2,2}\left( \omega ,Q\right)
+P^{3,3}\left( \omega ,Q\right)
\end{equation}%
because $P^{2,2}\left( \omega ,Q\right) $ and $P^{3,3}\left( \omega
,Q\right) $ involve operators that create or destroy internodal
electron-hole pairs.

If we sum over $k$ in the GRPA\ equation, we get%
\begin{eqnarray}
&&I\left( \omega +i\delta \right) \overline{P}\left( \omega ,Q\right)  \notag
\label{sumpp} \\
&&-\frac{1}{L_{z}}\sum_{k}\left( \overline{E}\left( k,Q\right) +\overline{%
\Sigma }\left( k\right) \right) \overline{P}_{k+Q,k}\left( \omega \right) 
\notag \\
&&-\frac{1}{L_{z}}\sum_{k}\overline{B}\left( k\right) \overline{H}\left(
Q\right) \overline{P}\left( \omega ,Q\right)  \notag \\
&&+\frac{1}{L_{z}}\sum_{k^{\prime }}\left[ \frac{1}{L_{z}}\sum_{k}\overline{B%
}\left( k\right) \overline{X}\left( k-k^{\prime }\right) \right] \overline{P}%
_{k^{\prime }+Q,k^{\prime }}\left( \omega \right)  \notag \\
&=&\frac{1}{L_{z}}\sum_{k}\overline{B}\left( k\right) .
\end{eqnarray}%
But, by the definition of the self-energy, we have the result%
\begin{equation}
\frac{1}{L_{z}}\sum_{k}\overline{B}\left( k\right) \overline{X}\left(
k-k^{\prime }\right) =\overline{\Sigma }\left( k^{\prime }\right) .
\end{equation}%
Thus, when performing a summation over $k,$ the self-energies in the
equation for $\overline{P}\left( \omega ,Q\right) $ are exactly canceled by
the vertex corrections due to the ladder diagrams. Moreover, since $e_{\pm
}\left( k\right) -e_{\pm }\left( k+Q\right) =\pm \hslash v_{F}Q$ does not
involve $k,$ the summation over $k$ gives directly (apart from a
multiplicative constant), the responses $\chi _{nn}\left( \omega ,Q\right) $
and $\chi _{jj}\left( \omega ,Q\right) .$ The same procedure cannot be
applied to the excitonic response since $e_{\pm }\left( k\right) -e_{\mp
}\left( k+Q\right) =\left( \mp 2k\mp Q\right) \hslash v_{F}$ which is not
just a function of $Q.$ This cancellation is an example of a Ward identity
and it occurs here because the noninteracting electronic dispersion is
linear and the current operator has the special form $j_{z,\tau }=-e\tau
v_{F}\sigma _{z}$ which is independent of $k.$ We are thus left, for the
components $P^{I,J}\left( \omega ,Q\right) ,$ with $I,J=1,4,$ (for $C=1$)
with the equation

\begin{eqnarray}
&&\left( 
\begin{array}{cc}
\begin{array}{c}
\omega +i\delta -v_{z}Q \\ 
-a_{+}\left( Q\right) H\left( Q\right)%
\end{array}
& -a_{+}\left( Q\right) H\left( Q\right) \\ 
-a_{-}\left( Q\right) H\left( Q\right) & 
\begin{array}{c}
\omega +i\delta +v_{z}Q \\ 
-a_{-}\left( Q\right) H\left( Q\right)%
\end{array}%
\end{array}%
\right)  \notag \\
&&\times \left( 
\begin{array}{cc}
P^{1,1}\left( \omega ,Q\right) & P^{1,4}\left( \omega ,Q\right) \\ 
P^{4,1}\left( \omega ,Q\right) & P^{4,4}\left( \omega ,Q\right)%
\end{array}%
\right)  \notag \\
&=&\left( 
\begin{array}{cc}
a_{+}\left( Q\right) & 0 \\ 
0 & a_{-}\left( Q\right)%
\end{array}%
\right) ,  \label{densicourant}
\end{eqnarray}%
where we have defined%
\begin{equation}
a_{\pm }\left( Q\right) =\frac{1}{2\pi \ell }\int_{-k_{c}\ell }^{+k_{c}\ell
}dk\ell \left[ \left\langle \rho _{\pm ,\pm }\left( k+Q\right) \right\rangle
-\left\langle \rho _{\pm ,\pm }\left( k\right) \right\rangle \right] .
\end{equation}%
Note that by symmetry, 
\begin{equation}
a_{+}\left( Q\right) =-a_{-}\left( Q\right) =a\left( Q\right) .
\end{equation}%
Solving Eq. (\ref{densicourant}) analytically, we arrive at the following
results for the density and current responses 
\begin{equation}
\chi _{nn}\left( \omega ,Q\right) =\frac{2e^{2}v_{F}a\left( Q\right) Q}{%
\left( \omega +i\delta \right) ^{2}-\frac{a\left( Q\right) }{Q\ell }\frac{%
v_{F}}{\ell }\frac{e^{2}}{\varepsilon _{0}\pi \hslash }-v_{F}^{2}Q^{2}}
\label{chinn}
\end{equation}%
and 
\begin{equation}
\chi _{jj}\left( \omega ,Q\right) =e^{2}v_{F}^{2}\frac{\frac{a^{2}\left(
Q\right) }{Q^{2}\ell ^{2}}\frac{2e^{2}}{\pi \varepsilon _{0}\hslash }%
+2v_{F}a\left( Q\right) Q}{\left( \omega +i\delta \right) ^{2}-\frac{a\left(
Q\right) }{Q\ell }\frac{v_{F}}{\ell }\frac{e^{2}}{\varepsilon _{0}\pi
\hslash }-v_{F}^{2}Q^{2}}.  \label{chijj}
\end{equation}%
Both functions have a single pole at the plasmon frequency $\omega _{p}$
given by

\begin{equation}
\omega _{p}=\sqrt{\frac{a\left( Q\right) }{Q\ell }\frac{v_{F}}{\ell }\frac{%
e^{2}}{\varepsilon _{0}\pi \hslash }+v_{F}^{2}Q^{2}}.  \label{plasmon2}
\end{equation}

In contrast, the proper responses $\widetilde{\chi }_{nn}\left( \omega
,Q\right) $ and $\widetilde{\chi }_{jj}\left( \omega ,Q\right) $ have a
single pole at the intranodal electron-hole excitation $e_{\pm }\left(
k\right) -e_{\pm }\left( k+Q\right) =\pm \hslash v_{F}Q$ of the
noninteracting electron gas. This pole is transformed into the plasmon mode
at a finite frequency by the Hartree term in the GRPA.

As we mentioned, the cancellation of the self-energy by the ladder diagrams
does not occur for the excitonic response and we are thus forced to compute
the full GRPA matrix equation numerically. This we do by discretizing the
wave vector $k$ in the exact same way as when solving the Hartree-Fock
equation for the single-particle Green's function. We obtain a matrix
equation that has the form%
\begin{equation}
\left[ I\left( \omega +i\delta \right) -\overline{\Upsilon }\left( k\right) %
\right] \overline{P}_{k+Q,k}\left( \omega \right) =\overline{B}\left(
k\right) ,
\end{equation}%
where $\overline{\Upsilon }\left( k\right) $ is a $4\left( 2N_{p}+1\right)
\times 4\left( 2N_{p}+1\right) $ matrix, with $2N_{p}+1$ the number of $k$
values used in the HFA calculation.

\section{RESPONSE FUNCTIONS IN THE INCOHERENT STATE}

In the incoherent but interacting ground state (doped or undoped), the
function 
\begin{equation}
a\left( Q\right) =\frac{Q}{2\pi }
\end{equation}%
so that the plasmon frequency is given exactly by%
\begin{equation}
\omega _{p}=\sqrt{\frac{e^{3}v_{F}B}{2\pi ^{2}\varepsilon _{0}\hslash ^{2}}%
+v_{F}^{2}Q^{2}}  \label{plasmon}
\end{equation}%
which is the well-known result\cite{Plasmon}. It does not depend on doping,
nor on self-energy and vertex corrections. The density and current responses
are exactly given by 
\begin{equation}
\chi _{nn}\left( \omega ,Q\right) =\frac{e^{2}v_{F}Q^{2}/\pi }{\left( \omega
+i\delta \right) ^{2}-\omega _{p}^{2}}
\end{equation}%
and%
\begin{equation}
\chi _{jj}\left( \omega ,Q\right) =e^{2}v_{F}^{2}\frac{\frac{e^{3}B}{2\pi
^{3}\varepsilon _{0}\hslash ^{2}}+\frac{v_{F}Q^{2}}{\pi }}{\left( \omega
+i\delta \right) ^{2}-\omega _{p}^{2}}
\end{equation}%
respectively. They have a single pole at the plasmon frequency. As $%
Q\rightarrow 0,$ the density response goes to zero but the current response
remains finite. The coherences enter these response functions and the
plasmon frequency only through the modification of the function $a\left(
Q\right) .$ Note that the continuum of excitations at $\omega =v_{F}Q$ has
been transformed into the plasmon mode by the Hartree term (bubble diagrams)
in the GRPA.

In the absence of interaction and for $Q=0,$ the excitonic response $\chi _{%
\text{exc}}^{\left( 0\right) }$ is given by 
\begin{eqnarray}
\func{Im}\left[ -\chi _{\text{exc}}^{\left( 0\right) }\left( \omega
,Q=0\right) \right] &=&\frac{1}{2v_{F}}\Theta \left( 2v_{F}k_{F}-\omega
\right)  \notag \\
&&-\frac{1}{2v_{F}}\Theta \left( 2v_{F}k_{F}+\omega \right)
\end{eqnarray}%
and there is only a continuum of electron-hole pair excitations. With
interaction but in the incoherent state where $\left\langle \rho
_{-,+}\left( k\right) \right\rangle =0$, the density and current responses
are uncoupled from the excitonic response $\chi _{\text{exc}}\left( \omega
,Q\right) $. Moreover, the bubble diagrams do not contribute to $\chi _{%
\text{exc}}\left( \omega ,Q\right) $ in this case. The excitonic response is
thus solution of the equation%
\begin{eqnarray}
&&\left[ \left( \omega +i\delta \right) -\left( \widetilde{e}_{\mp }\left(
k\right) -\widetilde{e}_{\pm }\left( k+Q\right) \right) /\hslash \right]
P_{k+Q,k}^{\pm }\left( \omega \right)  \notag \\
&&+\frac{e^{2}}{\varepsilon _{0}\hslash }b_{\pm }(k,Q)\frac{1}{L_{z}}%
\sum_{q}X\left( k-q\right) P_{q+Q,q}^{\pm }\left( \omega \right)  \notag \\
&=&b_{\pm }(k,Q),  \label{p223}
\end{eqnarray}%
where we have defined%
\begin{equation}
b_{\pm }(k,Q)=\left\langle \rho _{\pm ,\pm }\left( k+Q\right) \right\rangle
-\left\langle \rho _{\mp ,\mp }\left( k\right) \right\rangle
\end{equation}%
and 
\begin{equation}
\widetilde{e}_{\tau }\left( k\right) =e_{\tau }\left( k\right) +\Sigma
_{\tau }\left( k\right) .
\end{equation}%
The excitonic response in the GRPA\ is found by first solving the Eq. (\ref%
{p223}) where the upper(lower) sign is for $P^{22}\left( P^{33}\right) $ and
then summing over $k.$

\section{EXCITONIC RESPONSE IN THE COHERENT STATE WITH\ $C=1$}

Figure 9 shows the imaginary part of the excitonic and current response
functions for $Q\ell =0.15$ and $v_{F}/c=0.001$ for nodes with Chern number $%
C=1.$ The full(dashed) line is the GRPA(proper)\ response. Since, in the
current response, the ladder diagrams cancel the self-energies in Eq. (\ref%
{sumpp}), the proper responses $\widetilde{\chi }_{jj}\left( \omega
,Q\right) $ and $\widetilde{\chi }_{nn}\left( \omega ,Q\right) $ have only
one pole which is at a frequency $\omega =\left( e_{+}\left( k\right)
-e_{+}\left( k+Q\right) \right) /\hslash =v_{F}Q$ for $\omega \geq 0.$ In
the coherent phase, however, all response functions are coupled by the
internodal coherence and so this mode also appears in $\widetilde{\chi }_{%
\text{exc}}\left( \omega ,Q\right) $ (the first peak in the green dashed
line). The other peaks in $\widetilde{\chi }_{\text{exc}}\left( \omega
,Q\right) $ at energies $E_{n}$ (with $n=1,2,...$) are electron-hole bound
states (excitons). Their energy increases with $n$ until the energy of the
electron-hole internodal continuum whose onset is $E_{\text{conti}}\left(
Q\right) $ is reached (this onset is indicated by the vertical brown line in
Fig. 9). The bound state energies for $Q\rightarrow 0$ are approximately
given by $e_{B,n}=\left( E_{\text{conti}}\left( Q\right) -E_{1}\right)
/n^{x} $ where the exponent $x$ depends on the Fermi velocity. Because of
the vertex (ladder) corrections, the onset energy $E_{\text{conti}}\left(
Q\right) $ is slightly red shifted with respect to the Hartree-Fock gap $%
\Delta _{HF}\left( Q\right) =E_{+}\left( Q\right) -E_{-}\left( 0\right) $
indicated by the orange line in Fig. 9.

The gapless mode at $\omega =v_{F}Q$ is not a pole of $\widetilde{\chi }_{%
\text{exc}}\left( \omega ,Q\right) $ as calculated in the incoherent state
using Eq. (\ref{p223}). It appears in $\widetilde{\chi }_{\text{exc}}\left(
\omega ,Q\right) $ only in the coherent state. It is present in $\widetilde{%
\chi }_{jj}\left( \omega ,Q\right) $ and $\widetilde{\chi }_{nn}\left(
\omega ,Q\right) $ as an intraband single-particle excitation but since it
shows up in $\widetilde{\chi }_{\text{exc}}\left( \omega ,Q\right) $ as a
gapless mode, we assume that it is also the collective mode related to the
fluctuations of the global phase $\varphi $ of the complex order parameter $%
\left\langle \rho _{-+}\right\rangle .$ The series of excitonic bound states
could then be associated with fluctuations in the amplitude of the order
parameter. When the Hartree term is considered in calculating the GRPA\
response, this gapless mode is strongly renormalized and becomes gapped at
the plasmon frequency given by Eq. (\ref{plasmon2}). This frequency is
slightly modified by the internodal coherence from its value in the
incoherent phase which is given by Eq. (\ref{plasmon}). In contrast, the
frequency of the excitonic peaks (the bound states) are almost unchanged
when the Hartree term is switched on. In consequence, there is no gapless
(Goldstone) mode in the GRPA\ spectrum of collective excitations for $\chi _{%
\text{exc}}\left( \omega ,Q\right) $ but there is one in the proper
response. Similar results are obtained when the GRPA is applied to the study
of collective excitations in superconductors\cite{Anderson,Bardasis}.

\begin{figure}[tbph]
\centering\includegraphics[width = \linewidth]{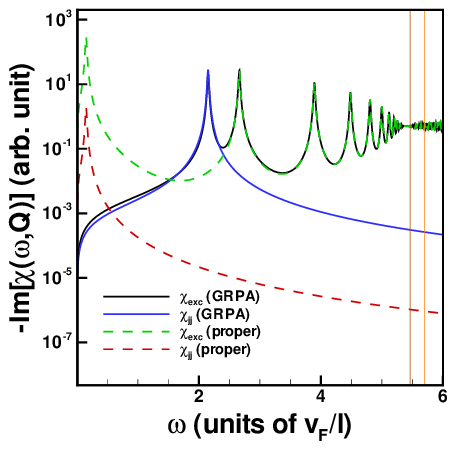}
\caption{Imaginary part of the excitonic and current response functions for $%
Q\ell =0.15$ and $v_{F}/c=0.001$ in the undoped coherent state. The full
lines are the GRPA\ results while the dashed line are the proper responses.
The vertical lines indicate the onset of the internodal continuum of
electron-hole excitations (brown line) and the Hartree-Fock gap at $k=0$
(orange line). The dielectric constant $\protect\varepsilon _{r}=1.$ }
\label{fig9}
\end{figure}

Figure 10 shows the GRPA\ response functions $\chi _{jj}\left( \omega
,Q\right) $ (blue line) and $\chi _{\text{exc}}\left( \omega ,Q\right) $
(black line) for $v_{F}/c=0.002$ and $Q\ell =0.015.$ Again the plasmon
appears as an extra pole in $\chi _{\text{exc}}\left( \omega ,Q\right) $
which is now in between two bound states. As $v_{F}/c$ increases and the
Hartree-Fock gap decreases, the plasmon pole eventually ends up in the
continuum of electron-hole internodal excitations.

\begin{figure}[tbph]
\centering\includegraphics[width = \linewidth]{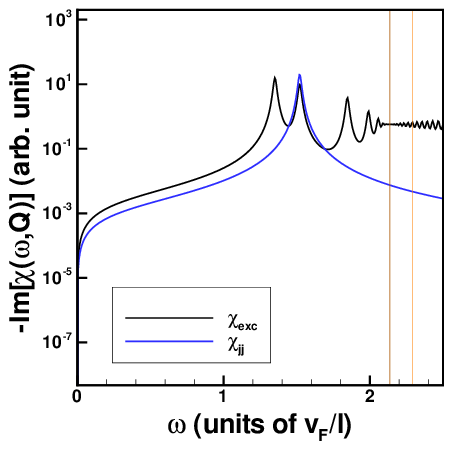}
\caption{Imaginary part of the GRPA\ excitonic (black curve) and current
(blue curve) response functions for $Q\ell =0.015$ and $v_{F}/c=0.002$ in
the undoped coherent phase. The vertical lines indicate the onset of the
internodal electron-hole continuum (brown) and the Hartree-Fock gap at $k=0$
(orange line). The dielectric constant $\protect\varepsilon _{r}=1.$ }
\label{fig10}
\end{figure}

Figure 11 shows $\chi _{jj}\left( \omega ,Q\right) $ (blue line) and $\chi _{%
\text{exc}}\left( \omega ,Q\right) $ (black line) for $v_{F}/c=0.001$ and $%
Q\ell =0.15$ in the presence of electron doping. The Fermi wave vector is $%
k_{F}\ell =0.1.$ The excitonic response shows two bound states before the
continuum of internodal electron-node excitations whose onset, indicated by
the vertical brown line, would be at $\omega =\left( E_{+}\left(
k_{F}\right) -E_{-}\left( k_{F}-Q\right) \right) /\hslash $ if vertex
corrections were neglected but is actually increased by them. As in the
undoped case, the excitonic response has an extra peak at the plasmon
frequency. The series of peaks at low frequency in $\chi _{\text{exc}}\left(
\omega ,Q\right) $ is the continuum of electron-hole excitations in the
upper Hartree-Fock band (see Fig. 4) which extends from $\omega =\left(
E_{+}\left( k_{F}\right) -E_{+}\left( k_{F}-Q\right) \right) /\hslash $ to $%
\omega =\left( E_{+}\left( k_{F}+Q\right) -E_{+}\left( k_{F}\right) \right)
/\hslash .$ As already noted, there is no continuum of excitations in $\chi
_{jj}\left( \omega ,Q\right) $ which has only the plasmon pole. For both
continua in $\chi _{\text{exc}}\left( \omega ,Q\right) $, the peaks are due
to our discretization of the wave vector $k$ which is needed to solve the
GRPA equations numerically.

\begin{figure}[tbph]
\centering\includegraphics[width = \linewidth]{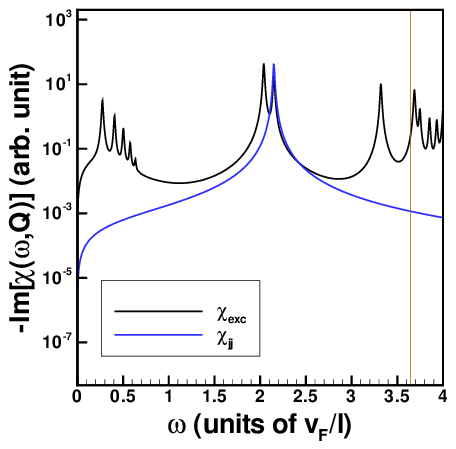}
\caption{Imaginary part of the GRPA\ excitonic and current response
functions for $Q\ell =0.15$ and $v_{F}/c=0.002$ in the doped coherent state
with Fermi wave vector $k_{F}\ell =0.1$. The vertical brown line indicates
the onset of the internode electron-hole continuum. The dielectric constant $%
\protect\varepsilon _{r}=1.$}
\label{fig11}
\end{figure}

\section{CONCLUSION}

We have studied the effect of the long-range Coulomb interaction on the
internodal coherence in a simple model of a two-node Weyl semimetal in the
extreme quantum limit. We have considered Weyl nodes with Chern number $%
C=1,2,3.$ As our numerical calculations show, it is possible to find values
of the Fermi velocity $v_{F},$ dielectric constant and the doping level $%
k_{F}$ where an excitonic condensate is possible and where the assumptions
made to justify our model are satisfied. Nevertheless, the order parameter
of the excitonic state decreases rapidly with the increase in all three
parameters.

At the mean-field level, the main effect of the internodal electron-hole
pairing is the opening of a gap in the chiral Landau levels making the
system insulating. Our calculation assumes that the two nodes are two
distinct systems and so this gap does not show up in the current or density
response function where only a plasmon pole is present whose frequency is
only slightly affected by the internodal coherence. If the two nodes are
viewed as one system, then the excitonic state manifests itself as a charge
density wave in real space. The gap and the sliding motion of the CDW should
then change the conductivity of the Weyl semimetal in several ways and cause
anomalous magnetoelectric transport effects as discussed extensively in the
literature (see Refs. \onlinecite{Yang2011}-\onlinecite{Curtis2023},%
\onlinecite{Gooth2019}). At the moment of writing this paper, a few papers
reported the experimental observation of an axionic CDW in the
quasi-one-dimensional Weyl semimetal (TaSe$_{4}$)$_{2}$I at zero magnetic
field\cite{Gooth2019,Shi2021}. These claims are, however, under an active
debate\cite{Sinchenko2022}.

In our calculation, the excitonic response function shows a series of
excitonic peaks in the gap opened by the internodal coherence. Their binding
energy decreases until the electron-hole (internodal) continuum is reached.
Because of the coupling between the different 16 response functions in the
GRPA, an extra gapped mode also appears as a peak in the excitonic response
function. In the proper excitonic response (ladder diagrams only), this peak
is at the frequency $\omega =\left( e_{\pm }\left( k\right) -e_{\pm }\left(
k+Q\right) \right) /\hslash =\pm v_{F}Q$ corresponding to a simple
noninteracting intranodal electron-hole excitation. This frequency is pushed
to the plasmon frequency when the full GRPA is computed by adding the bubble
diagrams.

The presence of a magnetic field modifies profoundly the excitonic state
with respect to its counterpart at zero magnetic field. The plasmon mode,
the density modulation and the Hartree-Fock gap (including its dependence on
doping), which are in principle measurable quantities, all depend on the
strength of the magnetic field. Equation (31) shows that the order parameter
(and so the phase diagram) of the excitonic phase can be obtained from the
amplitude of the density modulation. Another measurable observable of the
coherent state is its magneto-optical spectrum. The optical absorption is
related to the conductivity and to the proper part of the current response
function: $\sigma _{zz}\left( \omega \right) =i\widetilde{\chi }_{zz}\left(
\omega ,q=0\right) /\omega .$ At the level of approximation made in our
paper (keeping only the $n=0$ chiral levels), there is no signature of the
coherent state in absorption. More Landau levels need to be added to our
model in order to see how the inter-Landau-level transitions are modified
when coherence is present. This is not an easy task since more Landau levels
also means more types of coherence such as internodal and/or
inter-Landau-level. The size of the matrix that needs to be diagonalized in
the calculation of the current response becomes rapidly out of hand. We
leave this calculation for further work.

\begin{acknowledgments}
R. C\^{o}t\'{e} was supported by a grant from the Natural Sciences and
Engineering Research Council of Canada (NSERC) and S. F. Lopez by
scholarhips from NSERC and the Fonds de recherche du Qu\'{e}bec-Nature et
technologies (FRQNT). Computer time was provided by the Digital Research
Alliance of Canada.\smallskip
\end{acknowledgments}

\appendix{}

\section{HARTREE-FOCK FORMALISM FOR THE CALCULATION OF THE SINGLE-PARTICLE
GREEN'S FUNCTION IN A COHERENT STATE}

We give in this appendix a summary of the Hartree-Fock formalism for the
calculation of single-particle Green's function of a Weyl semimetal in a
magnetic field and in the coherent state. We present the general case where
an arbitrary number of Landau levels are considered and the Weyl nodes can
have a Chern number $C=1,2,3.$ We allow for all types of coherence:
internodal, inter-Landau-level and complete entanglement

We consider a simple model of a Weyl semimetal (WSM) with broken
time-reversal symmetry consisting of two nodes with Chern number $C$
centered at wave vectors $\mathbf{b}=-\tau b\widehat{\mathbf{z}}$ in
momentum space and with opposite chiralities $\tau =\pm 1.$The
noninteracting hamiltonian for each node, written in the basis of the two
bands that cross, is given by

\begin{equation}
h_{\tau }\left( \mathbf{k}\right) =\tau \hslash v_{F}\left( 
\begin{array}{cc}
k_{z} & \beta \left( k_{x}-ik_{y}\right) ^{C} \\ 
\beta \left( k_{x}+ik_{y}\right) ^{C} & -k_{z}%
\end{array}%
\right) ,
\end{equation}%
where $\beta $ is a material dependent anisotropy factor, $v_{F}$ is the
Fermi velocity and the wave vector $\mathbf{k}$ is restricted to a small
region around each node.

We consider the WSM to be in a magnetic field $\mathbf{B}=\mathbf{\nabla }%
\times \mathbf{A=}B\widehat{\mathbf{z}}.$ After making the Peierls
substitution $\mathbf{k\rightarrow k}+e\mathbf{A/\hslash }$ and working in
the Landau gauge $\mathbf{A}=\left( 0,Bx,0\right) ,$ we can write

\begin{equation}
h_{\tau }\left( \mathbf{k}\right) =\tau v_{F}\left( 
\begin{array}{cc}
\hslash k_{z} & \beta \left( \frac{\sqrt{2}\hslash }{\ell }a\right) ^{C} \\ 
\beta \left( \frac{\sqrt{2}\hslash }{\ell }a^{\dag }\right) ^{C} & -\hslash
k_{z}%
\end{array}%
\right) ,
\end{equation}%
where the ladder operators $a,a^{\dag },$ which obey the commutation
relation $\left[ a,a^{\dag }\right] =1,$ are defined by%
\begin{eqnarray}
a &=&\frac{\ell }{\sqrt{2}\hslash }\left( P_{x}-iP_{y}\right) , \\
a^{\dag } &=&\frac{\ell }{\sqrt{2}\hslash }\left( P_{x}+iP_{y}\right) ,
\end{eqnarray}%
and where $\mathbf{P}=\hslash \mathbf{k}+e\mathbf{A}$ with $-e$ the electron
charge.

We denote the different Landau levels by the set of indices $\left( n,s,\tau
\right) $ where $n=0,1,2,3,...$ and $s=\pm 1$ for the positive and negative
energy levels. For the chiral (linearly dispersing levels), $s=+1$ only. For
the nonchiral levels, the Landau level dispersions are given by (to simplify
the notation, we write $k$ instead of $k_{z}$ hereafter, the dispersion
being along $k_{z}$ only) 
\begin{eqnarray}
E_{n>0,s,k,\tau } &=&s\frac{\hslash v_{F}}{\ell }\sqrt{k^{2}\ell ^{2}+2n}, \\
E_{n>1,s,k,\tau } &=&s\frac{\hslash v_{F}}{\ell }\sqrt{k^{2}\ell ^{2}+4\beta
^{2}\left( \frac{\hslash }{\ell }\right) ^{2}n\left( n-1\right) },  \notag \\
E_{n>2,s,k,\tau } &=&s\frac{\hslash v_{F}}{\ell }\sqrt{k^{2}\ell ^{2}+8\beta
^{2}\left( \frac{\hslash }{\ell }\right) ^{4}n\left( n-1\right) \left(
n-2\right) },  \notag
\end{eqnarray}%
for $C=1,2,3$ respectively and the corresponding eigenvectors are given by%
\begin{equation}
w_{n,s,k,X,\tau }\left( \mathbf{r,}z\right) =\frac{1}{\sqrt{L_{z}}}\left( 
\begin{array}{c}
u_{n,s,k,\tau }h_{n-C,X}\left( \mathbf{r}_{\bot }\right) \\ 
v_{n\,,s,k,\tau }h_{n,X}\left( \mathbf{r}_{\bot }\right)%
\end{array}%
\right) e^{ikz},
\end{equation}%
where the factors 
\begin{equation}
\left( 
\begin{array}{c}
u_{n,s,k,\tau } \\ 
v_{n,s,k,\tau }%
\end{array}%
\right) =\frac{1}{\sqrt{2}}\left( 
\begin{array}{c}
s\tau \left( -i\right) ^{C}\sqrt{1+\frac{\hslash \tau v_{F}k}{E_{n,s,k,\tau }%
}} \\ 
\sqrt{1-\frac{\hslash \tau v_{F}k}{E_{n,s,k,\tau }}}%
\end{array}%
\right) .
\end{equation}%
In these equations, $X$ is the guiding-center index in the Landau gauge, $%
\mathbf{r}_{\bot }$ is a vector in the plane perpendicular to the magnetic
field and the factors and the functions $h_{n,X}\left( \mathbf{r}_{\bot
}\right) $ are defined by 
\begin{equation}
h_{n,X}\left( \mathbf{r}_{\bot }\right) =\varphi _{n}\left( x-X\right)
e^{-iXy/\ell ^{2}}/\sqrt{L_{y}},
\end{equation}%
where $\varphi _{n}\left( x\right) $ is a wave function of the
one-dimensional harmonic oscillator.

There are $C$ degenerate chiral levels at each node which we denote by the
integer $n$ ranging from $n=0$ to $n=C-1.$ They have the dispersion%
\begin{equation}
e_{\tau }\left( k\right) =-\tau \frac{\hslash v_{F}}{\ell }k\ell .
\end{equation}%
The corresponding eigenvectors are given by%
\begin{equation}
w_{n,s,k,X,\tau }\left( \mathbf{r}\right) =\frac{1}{\sqrt{L_{z}}}\left( 
\begin{array}{c}
0 \\ 
h_{n,X}\left( \mathbf{r}_{\bot }\right)%
\end{array}%
\right) e^{ikz}.
\end{equation}%
They are the same for both nodes. Since the Landau level energy is
independent of the quantum number $X,$ each state $\left( n,s,k,\tau \right) 
$ has degeneracy $N_{\varphi }=S/2\pi \ell ^{2}$ where $S=L_{x}L_{y}$ is the
area of the WSM\ perpendicular to the magnetic field. The volume of the WSM
is $L_{x}L_{y}L_{z}.$

We write the field operator for each node in the basis of these eigenvectors
so that 
\begin{eqnarray}
\Psi _{\tau }\left( \mathbf{r}\right) &=&\frac{1}{\sqrt{L_{z}}}%
\sum_{n,s,k,X}\left( 
\begin{array}{c}
u_{n,s,k,\tau }h_{n-C,X}\left( \mathbf{r}_{\bot }\right) \\ 
v_{n\,,s,k,\tau }h_{n,X}\left( \mathbf{r}_{\bot }\right)%
\end{array}%
\right) \\
&&\times e^{ikz}c_{n,s,k,X,\tau },  \notag
\end{eqnarray}%
where $c_{n,s,k,X,\tau }$ is the destruction operator for an electron in
state $\left( n,s,k,X,\tau \right) .$The many-body hamiltonian is then

\begin{eqnarray}
H &=&\sum_{\tau }\int d^{3}r\Psi _{\tau }^{\dagger }\left( \mathbf{r}\right)
h_{\tau }\left( \mathbf{r}\right) \Psi _{\tau }\left( \mathbf{r}\right) 
\notag \\
&&+\frac{1}{2}\sum_{\tau ,\tau ^{\prime }}\int d^{3}r\int d^{3}r^{\prime
}\Psi _{\tau }^{\dagger }\left( \mathbf{r}\right) \Psi _{\tau ^{\prime
}}^{\dagger }\left( \mathbf{r}^{\prime }\right)  \notag \\
&&\times V\left( \mathbf{r}-\mathbf{r}^{\prime }\right) \Psi _{\tau ^{\prime
}}\left( \mathbf{r}^{\prime }\right) \Psi _{\tau }\left( \mathbf{r}\right) ,
\end{eqnarray}%
where the Coulomb interaction (in S. I. units) 
\begin{equation}
V\left( \mathbf{r}\right) =\frac{1}{V}\sum_{\mathbf{q}}\frac{e^{2}}{%
\varepsilon _{r}\varepsilon _{0}\left\vert q_{\bot
}^{2}+q_{z}^{2}\right\vert }e^{i\mathbf{q}_{\bot }\cdot \mathbf{r}_{\bot
}}e^{iq_{z}z},
\end{equation}%
where the vector $\mathbf{q}_{\bot }=q_{x}\widehat{\mathbf{x}}+q_{y}\widehat{%
\mathbf{y}}$ and $\varepsilon _{r}$ is the dielectric constant of the WSM.
We remark that writing $e^{-i\left( \pm b+k\right) z}$ instead $e^{ikz}$ in
the field operators would make no change to $H$ so that the internodal
separation $2b$ does not enter our calculation. More precisely, we consider
the two nodes as distincts systems. Another sequence for the field
operators, namely $\Psi _{\tau }^{\dagger }\left( \mathbf{u}\right) \Psi
_{\tau ^{\prime }}^{\dagger }\left( \mathbf{u}^{\prime }\right) \Psi _{-\tau
^{\prime }}\left( \mathbf{u}^{\prime }\right) \Psi _{-\tau }\left( \mathbf{u}%
\right) $ with $\tau \neq \tau ^{\prime }$ could also be considered as it
conserves the number of particles in each node. However, it leads to Fourier
components of Coulomb interaction of the form $e^{2}/\left[ \varepsilon
_{r}\varepsilon _{0}\left( q_{\bot }^{2}+\left( q_{z}\pm 2b\right)
^{2}\right) \right] .$ We assume that $b$ is sufficiently large for these
terms to be negligeable in comparison with those that we keep.

\begin{widetext}

In the $\left( n,s,k,X,\tau \right) $ basis, the hamiltonian is given by 
\begin{eqnarray}
H &=&\sum_{t,k,X,\tau }E_{t,k,\tau }c_{t,k,X,\tau }^{\dag }c_{t,k,X,\tau } \\
&&+\frac{1}{2L_{z}S}\sum_{\mathbf{q}}\frac{e^{2}}{\varepsilon
_{0}\varepsilon _{r}\left( q_{\bot }^{2}+q_{z}^{2}\right) }\sum_{\tau ,\tau
^{\prime }}\sum_{k_{1},k_{2}}\sum_{t_{1},...,t_{4}}\sum_{X_{1},...X_{4}} 
\notag \\
&&\times \int d^{2}r_{\bot }w_{t_{1},k_{1},X_{1},\tau }^{\dag }\left( 
\mathbf{r}_{\bot }\right) e^{i\mathbf{q}_{\bot }\cdot \mathbf{r}_{\bot
}}w_{t_{4},k_{1}+q_{z},X_{4},\tau }\left( \mathbf{r}_{\bot }\right) \int
d^{2}r_{\bot }^{\prime }w_{t_{2},k_{2},X_{2},\tau ^{\prime }}^{\dag }\left( 
\mathbf{r}_{\bot }^{\prime }\right) e^{-i\mathbf{q}_{\bot }\cdot \mathbf{r}%
_{\bot }^{\prime }}w_{t_{3},k_{2}-q_{z},X_{3},\tau ^{\prime }}\left( \mathbf{%
r}_{\bot }^{\prime }\right)   \notag \\
&&\times c_{t_{1},k_{1},X_{1},\tau }^{\dag }c_{t_{2},k_{2},X_{2},\tau
^{\prime }}^{\dag }c_{t_{3},k_{2}-q_{z},X_{3},\tau ^{\prime
}}c_{t_{4},k_{1}+q_{z},X_{4},\tau },  \notag
\end{eqnarray}%
where we have defined the super-index $t=\left( n,s\right) $ to lighten the
notation. The matrix elements%
\begin{equation}
\int d^{2}r_{\bot }w_{t_{1},k_{1},X_{1},\tau }^{\dag }\left( \mathbf{r}%
_{\bot }\right) e^{\pm i\mathbf{q}_{\bot }\cdot \mathbf{r}_{\bot
}}w_{t_{2},k_{2},X_{2},\tau }\left( \mathbf{r}_{\bot }\right) =e^{\pm \frac{i%
}{2}q_{x}\left( X_{1}+X_{2}\right) }\Lambda
_{t_{1},k_{1};t_{2},k_{2}}^{\left( \tau \right) }\left( \pm \mathbf{q}_{\bot
}\right) \delta _{X_{1},X_{2}\mp q_{y}\ell ^{2}},
\end{equation}%
where we have defined%
\begin{eqnarray}
\Lambda _{t_{1},k_{1};t_{2},k_{2}}^{\left( \tau \right) }\left( \mathbf{q}%
_{\bot }\right)  &=&u_{t_{1},k_{1},\tau }^{\ast }u_{t_{2},k_{2},\tau
}F_{n_{1}-1,n_{2}-1}\left( \mathbf{q}_{\bot }\right) +v_{t_{1},k_{1},\tau
}^{\ast }v_{t_{2},k_{2},\tau }F_{n_{1},n_{2}}\left( \mathbf{q}_{\bot
}\right) ,\text{ if }n_{1},n_{2}\geq C,  \label{a3} \\
\Lambda _{t_{1},k_{1};t_{2},k_{2}}^{\left( \tau \right) }\left( \mathbf{q}%
_{\bot }\right)  &=&v_{t_{1},k_{1},\tau }^{\ast }F_{n_{1},n_{2}}\left( 
\mathbf{q}_{\bot }\right) ,\text{ if }n_{1}\geq C\text{ and }n_{2}<C,  \notag
\\
\Lambda _{t_{1},k_{1};t_{2},k_{2}}^{\left( \tau \right) }\left( \mathbf{q}%
_{\bot }\right)  &=&v_{t_{2},k_{2},\tau }F_{n_{1},n_{2}}\left( \mathbf{q}%
_{\bot }\right) ,\text{ if }n_{2}\geq C\text{ and }n_{1}<C,  \notag \\
\Lambda _{t_{1},k_{1};t_{2},k_{2}}^{\left( \tau \right) }\left( \mathbf{q}%
_{\bot }\right)  &=&F_{n_{1},n_{2}}\left( \mathbf{q}_{\bot }\right) ,\text{
if }n_{1},n_{2}<C,  \notag
\end{eqnarray}%
and the function 
\begin{equation}
F_{n_{1},n_{2}}\left( \mathbf{q}_{\bot }\right) =\sqrt{\frac{Min\left(
n_{1},n_{2}\right) !}{Max\left( n_{1},n_{2}\right) !}}\left( \frac{iq_{\bot
}\ell }{\sqrt{2}}\right) ^{\left\vert n_{1}-n_{2}\right\vert }e^{-i\left(
n_{1}-n_{2}\right) \theta }L_{Min\left( n_{1},n_{2}\right) }^{\left\vert
n_{1}-n_{2}\right\vert }\left( \frac{q_{\bot }^{2}\ell ^{2}}{2}\right) e^{-%
\frac{q_{\bot }^{2}\ell ^{2}}{4}},
\end{equation}%
where $\theta $ is the angle between the vector $\mathbf{q}_{\bot }$ and the 
$x$ axis and $L_{n}^{m}\left( x\right) $ is a generalized Laguerre
polynomial.

At this point, we define the operators%
\begin{equation}
\rho _{t,k,\tau ;t^{\prime },k^{\prime },\tau ^{\prime }}\left( \mathbf{q}%
_{\bot }\right) \equiv \frac{1}{N_{\varphi }}\sum_{X,X^{\prime }}e^{-\frac{i%
}{2}q_{x}\left( X+X^{\prime }\right) }\delta _{X,X^{\prime }+q_{y}\ell
^{2}}c_{t,k,X,\tau }^{\dagger }c_{t^{\prime },k^{\prime },X^{\prime },\tau
^{\prime }}.
\end{equation}%
The set of averages $\left\{ \left\langle \rho _{t,k,\tau ;t^{\prime
},k^{\prime },\tau ^{\prime }}\left( \mathbf{q}_{\bot }\right) \right\rangle
\right\} $ define any phase of the electron gas in the WSM whether uniform
or modulated spatially and with any type of coherences: internodal,
inter-Landau-level or complete entanglement.

After making the Hartree-Fock pairing of the operators in the interacting
hamiltonian, we get 
\begin{eqnarray}
H_{HF} &=&N_{\varphi }\sum_{t,k,\tau }E_{t,k,\tau }\rho _{t,k,\tau ;t,k,\tau
}\left( 0\right) \\
&&+\frac{N_{\varphi }^{2}}{L_{z}}\sum_{\mathbf{q}}\sum_{\tau ,\tau ^{\prime
}}\sum_{k_{1},k_{2}}%
\sum_{t_{1},...,t_{4}}H_{t_{1},k_{1};t_{4},k_{1}+q_{z};t_{2},k_{2};t_{3},k_{2}-q_{z}}^{\left( \tau ,\tau ^{\prime }\right) }\left( 
\mathbf{q}\right) \left\langle \rho _{t_{1},k_{1},\tau
;t_{4},k_{1}+q_{z},\tau }\left( -\mathbf{q}_{\bot }\right) \right\rangle
\rho _{t_{2},k_{2},\tau ^{\prime };t_{3},k_{2}-q_{z},\tau ^{\prime }}\left( 
\mathbf{q}_{\bot }\right)  \notag \\
&&-\frac{N_{\varphi }}{L_{z}}\sum_{\mathbf{q}}\sum_{\tau ,\tau ^{\prime
}}\sum_{k_{1},k_{2}}%
\sum_{t_{1},...,t_{4}}X_{t_{1},k_{1};t_{4},k_{1}+q_{z};t_{2},k_{2};t_{3},k_{2}-q_{z}}^{\left( \tau ,\tau ^{\prime }\right) }\left( 
\mathbf{q}\right) \left\langle \rho _{t_{1},k_{1},\tau
;t_{3},k_{2}-q_{z},\tau ^{\prime }}\left( -\mathbf{q}_{\bot }\right)
\right\rangle \rho _{t_{2},k_{2},\tau ^{\prime };t_{4},k_{1}+q_{z},\tau
}\left( \mathbf{q}_{\bot }\right) ,  \notag
\end{eqnarray}%
where $\mathbf{t}_{\bot }=t_{x}\widehat{\mathbf{x}}+t_{y}\widehat{\mathbf{y}}
$ and the Hartree and Fock interactions are given by 
\begin{eqnarray}
H_{t_{1},k_{1};t_{2},k_{2};t_{3},k_{3};t_{4},k_{4}}^{\left( \tau ,\tau
^{\prime }\right) }\left( \mathbf{q}\right) &=&\frac{1}{S}\Lambda
_{t_{1},k_{1};t_{2}k_{2}}^{\left( \tau \right) }\left( \mathbf{q}_{\bot
}\right) V\left( \mathbf{q}\right) \Lambda _{t_{3}k_{3};t_{4},k_{4}}^{\left(
\tau ^{\prime }\right) }\left( -\mathbf{q}_{\bot }\right) , \\
X_{t_{1},k_{1};t_{2},k_{2};t_{3},k_{3};t_{4},k_{4}}^{\left( \tau ,\tau
^{\prime }\right) }\left( \mathbf{q}\right) &=&\frac{1}{S}\sum_{\mathbf{t}%
_{\bot }}e^{-i\mathbf{t}_{\bot }\times \mathbf{q}_{\bot }\ell ^{2}}\Lambda
_{t_{1},k_{1};t_{2},k_{2}}^{\left( \tau \right) }\left( \mathbf{t}_{\bot
}\right) V\left( \mathbf{t}_{\bot },q_{z}\right) \Lambda
_{t_{3},k_{3};t_{4},k_{4}}^{\left( \tau ^{\prime }\right) }\left( -\mathbf{t}%
_{\bot }\right) .
\end{eqnarray}

To compute the $\left\langle \rho _{t,k,\tau ;t^{\prime },k^{\prime },\tau
^{\prime }}\left( \mathbf{q}_{\bot }\right) \right\rangle ^{\prime }$s, we
define the single-particle Matsubara Green's function%
\begin{equation}
G_{t,k,\tau ;t^{\prime },k^{\prime },\tau ^{\prime }}\left( \mathbf{q}_{\bot
};\tau _{0}\right) =\frac{1}{N_{\varphi }}\sum_{X,X^{\prime }}e^{-\frac{i}{2}%
q_{x}\left( X+X^{\prime }\right) }\delta _{X,X^{\prime }-q_{y}\ell
^{2}}G_{t,k,\tau ;t^{\prime },k^{\prime },\tau ^{\prime }}\left( X,X^{\prime
};\tau _{0}\right) ,
\end{equation}%
where the imaginary-time Green's function is defined as%
\begin{equation}
G_{t,k,\tau ;t^{\prime },k^{\prime },\tau ^{\prime }}\left( X,X^{\prime
};\tau _{0}\right) =-\left\langle T_{\tau _{0}}c_{t,k,X,\tau }\left( \tau
_{0}\right) c_{t^{\prime },k^{\prime },X^{\prime },\tau ^{\prime }}^{\dagger
}\left( 0\right) \right\rangle ,
\end{equation}%
with $T_{\tau _{0}}$ the imaginary-time ordering operator and $\tau _{0}$
the imaginary time (not to be confused with the node index). When $\tau
_{0}=0^{-},$%
\begin{equation}
G_{t,k,\tau ;t^{\prime },k^{\prime },\tau ^{\prime }}\left( \mathbf{q}_{\bot
};\tau _{0}=0^{-}\right) =\left\langle \rho _{t^{\prime },k^{\prime },\tau
^{\prime };t,k,\tau }\left( \mathbf{q}_{\bot }\right) \right\rangle .
\end{equation}

Now, using the Fourier transform%
\begin{equation}
G_{t,k,\tau ;t^{\prime },k^{\prime },\tau ^{\prime }}\left( \mathbf{q}_{\bot
}\mathbf{,}i\omega _{m}\right) =\int_{0}^{\beta \hslash }d\tau
_{0}e^{i\omega _{m}\tau _{0}}G_{t,k,\tau ;t^{\prime },k^{\prime },\tau
^{\prime }}\left( \mathbf{q}_{\bot },\tau _{0}\right) ,
\end{equation}%
with the Matsubara fermionic frequencies 
\begin{equation}
\omega _{m}=\frac{\left( 2m+1\right) \pi }{\beta \hslash },\qquad m=0,\pm
1,\pm 2,...
\end{equation}%
and $\beta =1/k_{B}T$ with $T$ the temperature and $k_{B}$ the Boltzmann
constant, we finally obtain the seeked averages by performing the Matsubara
frequency sum%
\begin{equation}
\left\langle \rho _{t^{\prime },k^{\prime },\tau ^{\prime };t,k,\tau }\left( 
\mathbf{q}_{\bot }\right) \right\rangle =\frac{1}{\beta \hslash }%
\sum_{i\omega _{m}}e^{-i\omega _{m}0^{-}}G_{t,k,\tau ;t^{\prime },k^{\prime
},\tau ^{\prime }}\left( \mathbf{q}_{\bot },i\omega _{m}\right) .
\end{equation}

It remains to derive the Hartree-Fock equation for the Green's function $%
G_{t,k,\tau ;t^{\prime },k^{\prime },\tau ^{\prime }}\left( \mathbf{q}_{\bot
},i\omega _{m}\right) .$ This is done by using the Heisenberg equation of
motion%
\begin{equation}
\hslash \frac{\partial }{\partial \tau _{0}}\left( \ldots \right) =\left[
H-\mu N_{e},\left( \ldots \right) \right] ,
\end{equation}%
where $\mu $ is the chemical potential and $N_{e}$ the electron number
operator. After a long calculation, we get%
\begin{align}
& \left[ i\omega _{n}-\frac{1}{\hslash }\left( E_{t,k,\tau }-\mu \right) %
\right] G_{t,k,\tau ;t^{\prime },k^{\prime },\tau ^{\prime }}\left( \mathbf{q%
}_{\bot },i\omega _{n}\right)   \notag \\
& =\delta _{\tau ,\tau ^{\prime }}\delta _{t,t^{\prime }}\delta
_{k,k^{\prime }}\delta _{\mathbf{q}_{\bot },0}  \notag \\
& +\frac{N_{\varphi }}{\hslash L_{z}}\sum_{\mathbf{q}^{\prime }}\sum_{\tau
^{\prime \prime
}}\sum_{k_{1}}\sum_{t_{1},t_{3},t_{4}}H_{t_{1},k_{1};t_{4},k_{1}+q_{z}^{%
\prime };t,k;t_{3},k-q_{z}^{\prime }}^{\left( \tau ^{\prime \prime },\tau
\right) }\left( \mathbf{q}_{\bot }^{\prime }-\mathbf{q}_{\bot
},q_{z}^{\prime }\right) \left\langle \rho _{t_{1},k_{1},\tau ^{\prime
\prime };t_{4},k_{1}+q_{z}^{\prime },\tau ^{\prime \prime }}\left( \mathbf{q}%
_{\bot }-\mathbf{q}_{\bot }^{\prime }\right) \right\rangle   \notag \\
& \times e^{-\frac{i}{2}\left( \mathbf{q}_{\bot }\times \mathbf{q}_{\bot
}^{\prime }\right) \ell ^{2}\cdot \widehat{\mathbf{z}}}G_{t_{3},k-q_{z}^{%
\prime },\tau ;t^{\prime },k^{\prime },\tau ^{\prime }}\left( \mathbf{q}%
_{\bot }^{\prime };\omega _{n}\right)   \notag \\
& -\frac{1}{\hslash L_{z}}\sum_{\mathbf{q}^{\prime }}\sum_{\tau ^{\prime
\prime
}}\sum_{k_{1}}\sum_{t_{1},t_{3},t_{4}}X_{t_{1},k_{1};t_{4},k_{1}+q_{z}^{%
\prime };t,k;t_{3},k-q_{z}^{\prime }}^{\left( \tau ^{\prime \prime },\tau
\right) }\left( \mathbf{q}_{\bot }^{\prime }-\mathbf{q}_{\bot
},q_{z}^{\prime }\right) \left\langle \rho _{t_{1},k_{1},\tau ^{\prime
\prime };t_{3},k-q_{z}^{\prime },\tau }\left( \mathbf{q}_{\bot }-\mathbf{q}%
_{\bot }^{\prime }\right) \right\rangle   \notag \\
& \times e^{-\frac{i}{2}\left( \mathbf{q}_{\bot }\times \mathbf{q}_{\bot
}^{\prime }\right) \ell ^{2}\cdot \widehat{\mathbf{z}}%
}G_{t_{4},k_{1}+q_{z}^{\prime },\tau ^{\prime \prime };t^{\prime },k^{\prime
},\tau ^{\prime }}\left( \mathbf{q}_{\bot }^{\prime };\omega _{n}\right) .
\end{align}

The average of the electronic density is given by%
\begin{eqnarray}
\left\langle n_{e}\left( \mathbf{q}\right) \right\rangle &=&\sum_{\tau }\int
d^{3}r\left\langle \Psi _{\tau }^{\dag }\left( \mathbf{r}\right) e^{-i%
\mathbf{q}\cdot \mathbf{r}}\Psi _{\tau }\left( \mathbf{r}\right)
\right\rangle \\
&=&N_{\varphi }\sum_{\tau ,t,t^{\prime },k}\Lambda _{t,k;t^{\prime
},k+q_{z}}^{\left( \tau \right) }\left( -\mathbf{q}_{\bot }\right)
\left\langle \rho _{t,k,\tau ;t^{\prime },k+q_{z},\tau }\left( \mathbf{q}%
_{\bot }\right) \right\rangle  \notag
\end{eqnarray}%
an so in the particular case where the electron gas is not modulated
spatially, we must have $\left\langle n_{e}\left( \mathbf{q}\right)
\right\rangle \neq 0$ for $q=0$ only which implies that%
\begin{equation}
\left\langle \rho _{t^{\prime },k^{\prime },\tau ^{\prime };t,k,\tau }\left( 
\mathbf{q}_{\bot }\right) \right\rangle =\left\langle \rho _{t^{\prime
},k,\tau ^{\prime };t,k,\tau }\left( 0\right) \right\rangle \delta
_{k,k^{\prime }}\delta _{\mathbf{q}_{\bot },0}.
\end{equation}

This condition simplifies the hamiltonian which becomes%
\begin{eqnarray}
H_{HF} &=&N_{\varphi }\sum_{t,k,\tau }E_{t,k,\tau }\rho _{t,t}^{\left( \tau
,\tau \right) }\left( k\right) \\
&&-\frac{N_{\varphi }}{L_{z}}\sum_{\tau ,\tau ^{\prime
}}\sum_{k_{1},k_{2}}%
\sum_{t_{1},...,t_{4}}X_{t_{1},k_{1};t_{4},k_{2};t_{2},k_{2};t_{3},k_{1}}^{%
\left( \tau ,\tau ^{\prime }\right) }\left( q_{z}=k_{2}-k_{1}\right)
\left\langle \rho _{t_{1},t_{3}}^{\left( \tau ,\tau ^{\prime }\right)
}\left( k_{1}\right) \right\rangle \rho _{t_{2},t_{4}}^{\left( \tau ^{\prime
},\tau \right) }\left( k_{2}\right)  \notag
\end{eqnarray}%
and the equation for the single-particle Green's function also simplifies to%
\begin{equation}
\left[ i\omega _{n}-\frac{1}{\hslash }\left( E_{t,\tau }\left( k\right) -\mu
\right) \right] G_{t,t^{\prime }}^{\left( \tau ,\tau ^{\prime }\right)
}\left( k,i\omega _{n}\right) -\frac{1}{\hslash }\sum_{\tau ^{\prime \prime
},t^{\prime \prime }}\Sigma _{t,t^{\prime \prime }}^{\left( \tau ,\tau
^{\prime \prime }\right) }\left( k\right) G_{t^{\prime \prime },t^{\prime
}}^{\left( \tau ^{\prime \prime },\tau ^{\prime }\right) }\left( k,i\omega
_{n}\right) =\delta _{\tau ,\tau ^{\prime }}\delta _{t,t^{\prime }},
\label{a100}
\end{equation}%
where we have defined the Fock self-energy%
\begin{equation}
\Sigma _{t,t^{\prime }}^{\left( \tau ,\tau ^{\prime }\right) }\left(
k\right) =-\frac{1}{L_{z}}\sum_{k_{1}}\sum_{t_{1},t_{2}}X_{t_{1},k_{1};t^{%
\prime },k;t,k;t_{2},k_{1}}^{\left( \tau ^{\prime },\tau \right) }\left(
0,k-k_{1}\right) \left\langle \rho _{t_{1},t_{2}}^{\left( \tau ^{\prime
},\tau \right) }\left( k_{1}\right) \right\rangle
\end{equation}%
and simplified the notation to%
\begin{eqnarray}
G_{t,t^{\prime }}^{\left( \tau ,\tau ^{\prime }\right) }\left( k,i\omega
_{n}\right) &=&G_{t,k,\tau ;t^{\prime },k,\tau ^{\prime }}\left( \mathbf{q}%
_{\bot }=0,i\omega _{n}\right) , \\
\left\langle \rho _{t,t^{\prime }}^{\left( \tau ,\tau ^{\prime }\right)
}\left( k\right) \right\rangle &=&\left\langle \rho _{t,k,\tau ;t^{\prime
},k,\tau ^{\prime }}\left( \mathbf{q}_{\bot }=0\right) \right\rangle .
\end{eqnarray}%
In a uniform state, the Hartree term is cancelled by the positive ionic
background of the WSM and so there is no Hartree self-energy. The
Hartree-Fock equation of motion for $G_{t,t^{\prime }}^{\left( \tau ,\tau
^{\prime }\right) }\left( k,i\omega _{n}\right) $ is a self-consistent
equation since the self-energy contains the very averages that we want to
compute.

Defining the super-indices $I,J,K=\left( t,\tau \right) =1,2,3,..,N$ where $%
N $ is the total number of levels considered, eq. (\ref{a100}) can be
written as%
\begin{equation}
\sum_{K}\left[ i\omega _{n}\delta _{I,K}-F_{I,K}\left( k\right) \right]
G_{K,J}\left( k,i\omega _{n}\right) =\delta _{I,J},  \label{eqgreen}
\end{equation}%
where the matrix 
\begin{equation}
F_{I,J}\left( k\right) =\frac{1}{\hslash }\left[ \left( E_{t,\tau }\left(
k\right) -\mu \right) \delta _{I,J}-\Sigma _{I,J}\left( k\right) \right] .
\end{equation}

Because of the symmetry relations%
\begin{eqnarray}
\left\langle \rho _{I,J}\left( k\right) \right\rangle &=&\left\langle \rho
_{J,I}\left( k\right) \right\rangle ^{\ast }, \\
F_{n,n^{\prime }}\left( \mathbf{q}_{\bot }\right) &=&\left[ F_{n^{\prime
},n}\left( -\mathbf{q}_{\bot }\right) \right] ^{\ast }, \\
\Lambda _{t,k;t^{\prime }k^{\prime }}^{\left( \tau \right) }\left( \mathbf{q}%
_{\bot }\right) &=&\left[ \Lambda _{t^{\prime },k^{\prime };t,k}^{\left(
\tau \right) }\left( -\mathbf{q}_{\bot }\right) \right] ^{\ast },
\end{eqnarray}%
it follows that%
\begin{eqnarray}
H_{t_{1},k_{1};t_{2},k_{2};t_{3},k_{3};t_{4},k_{4}}^{\left( \tau ,\tau
^{\prime }\right) }\left( \mathbf{q}\right) &=&\left[
H_{t_{4},k_{4};t_{3},k_{3};t_{2},k_{2};t_{1},k_{1}}^{\left( \tau ^{\prime
},\tau \right) }\left( -\mathbf{q}\right) \right] ^{\ast }, \\
X_{t_{1},k_{1};t_{2},k_{2};t_{3},k_{3};t_{4},k_{4}}^{\left( \tau ,\tau
^{\prime }\right) }\left( \mathbf{q}\right) &=&\left[
X_{t_{4},k_{4};t_{3},k_{3};t_{2},k_{2};t_{1},k_{1}}^{\left( \tau ^{\prime
},\tau \right) }\left( -\mathbf{q}\right) \right] ^{\ast }  \notag
\end{eqnarray}%
and for the self-energies 
\begin{equation}
\Sigma _{t,t^{\prime }}^{\left( \tau ,\tau ^{\prime }\right) }\left(
k\right) =\left[ \Sigma _{t^{\prime },t}^{\left( \tau ^{\prime },\tau
\right) }\left( k\right) \right] ^{\ast }.
\end{equation}%
Thus, $F_{I,J}\left( k\right) $ is an hermitian matrix that can be
diagonalized by a unitary transformation. In matrix form,%
\begin{equation}
F\left( k\right) =U\left( k\right) D\left( k\right) U^{\dag }\left( k\right)
,
\end{equation}%
where $U\left( k\right) $ is the matrix of the eigenvectors of $F\left(
k\right) $ and $D\left( k\right) $ the diagonal matrix of its real
eigenvalues $d_{m}\left( k\right) $ where $m=1,2,...,N.$ The Green's
functions are given by%
\begin{equation}
G_{I,J}\left( k,i\omega _{n}\right) =\sum_{m=1}^{N}\frac{U_{I,m}\left(
k\right) \left( U^{\dag }\left( k\right) \right) _{m,J}}{i\omega _{n}+\mu
/\hslash -d_{m}\left( k\right) }.
\end{equation}%
Performing the Matsubara frequency sum, we get, at $T=0$ K, that the
ground-state averages are given by%
\begin{equation}
\left\langle \rho _{J,I}\left( k\right) \right\rangle
=\sum_{m=1}^{N}U_{I,m}\left( k\right) \left[ U^{\dag }\left( k\right) \right]
_{m,J}\Theta \left( e_{F}-d_{m}\left( k\right) \right) ,
\end{equation}%
where $e_{F}$ is the Fermi level wich is determined by the relation%
\begin{equation}
\sum_{I}\sum_{k}\left\langle \rho _{I,I}\left( k\right) \right\rangle
=\sum_{k}\sum_{m=1}^{N}\Theta \left( e_{F}-d_{m}\left( k\right) \right)
=N_{e},
\end{equation}%
where $N_{e}$ is the number of electrons in the two nodes.

In the approximation where we consider only the chiral levels in the Hilbert
space and where the state is uniform spatially, we have the simplification%
\begin{equation}
\Lambda _{t_{1},k_{1};t_{2},k_{2}}^{\left( \tau \right) }\left( \mathbf{q}%
_{\bot }\right) =F_{n_{1},n_{2}}\left( \mathbf{q}_{\bot }\right) ,
\end{equation}%
with the integers $n_{1},n_{2}$ ranging from $0$ to $C-1$ and with $%
t_{1}=n_{1},t_{2}=n_{2}$ since $s_{1}=s_{2}=1.$ The interactions are then
given by%
\begin{equation}
X_{n_{1},n_{2},n_{3},n_{4}}^{\left( \tau ,\tau ^{\prime }\right) }\left( 
\mathbf{q}_{\bot }=0,q_{z}\right) =\frac{e^{2}}{\varepsilon _{0}\varepsilon
_{r}}\int_{0}^{+\infty }\frac{dt_{\bot }\ell }{\left( 2\pi \right) ^{2}}%
\frac{t_{\bot }\ell }{t_{\bot }^{2}\ell ^{2}+q_{z}^{2}\ell ^{2}}%
\int_{0}^{2\pi }d\theta F_{n_{1},n_{2}}\left( \mathbf{t}_{\bot }\ell \right)
F_{n_{3},n_{4}}\left( -\mathbf{t}_{\bot }\ell \right) ,  \label{a1}
\end{equation}%
where $\theta $ is the angle between the vector $\mathbf{t}_{\bot }$ and the 
$x$ axis. They do not depend on the chirality index nor on the vectors $%
k_{1},...,k_{4}.$ The angular part is $\int_{0}^{2\pi }d\theta e^{-i\left(
n_{1}-n_{2}+n_{3}-n_{4}\right) \theta }$ and is finite only if $%
n_{1}-n_{2}+n_{3}-n_{4}=0.$ The nonzero interactions are thus given by%
\begin{eqnarray}
X_{n_{1},n_{2},n_{3},n_{4}}\left( x\right)  &=&\frac{e^{2}}{2\pi \varepsilon
_{0}\varepsilon _{r}}\sqrt{\frac{Min\left( n_{1},n_{2}\right) !}{Max\left(
n_{1},n_{2}\right) !}}\sqrt{\frac{Min\left( n_{3},n_{4}\right) !}{Max\left(
n_{3},n_{4}\right) !}}\int_{0}^{+\infty }dy\frac{y}{y^{2}+x^{2}}\left( \frac{%
y}{\sqrt{2}}\right) ^{\left\vert n_{1}-n_{2}\right\vert +\left\vert
n_{3}-n_{4}\right\vert }  \label{a4} \\
&&\times i^{\left\vert n_{1}-n_{2}\right\vert -\left\vert
n_{3}-n_{4}\right\vert }L_{Min\left( n_{1},n_{2}\right) }^{\left\vert
n_{1}-n_{2}\right\vert }\left( \frac{y^{2}}{2}\right) L_{Min\left(
n_{3},n_{4}\right) }^{\left\vert n_{3}-n_{4}\right\vert }\left( \frac{y^{2}}{%
2}\right) e^{-\frac{y^{2}}{2}}.  \notag
\end{eqnarray}%
They have the form $X_{n_{1},n_{2},n_{2},n_{1}}$ or $%
X_{n_{1},n_{1},n_{2},n_{2}}$ for $C=1,2.$\textit{\ }For $C=3,$one must add
to these terms the four terms $%
X_{2,1,0,1},X_{1,2,1,0},X_{0,1,2,1},X_{1,0,1,2}.$ Thus there are $1,6,$ and $%
19$ nonzero interactions for $C=1,2,3$ respectively. They are however not
all different as shown in appendix B.

\end{widetext}

\section{FOCK INTERACTIONS FOR $C=1,2,3$}

The nonzero Fock interactions, defined in Eq. (\ref{a4}), are given by ($%
x=k\ell $ and $\alpha =e^{2}/2\pi \varepsilon _{0}\varepsilon _{r}$):

\begin{widetext}

\subsection{ $C=1$}

\begin{equation}
X_{0,0,0,0}\left( x\right) =\alpha \frac{1}{2}e^{\frac{x^{2}}{2}}\Gamma
\left( 0,\frac{x^{2}}{2}\right) .
\end{equation}

\subsection{ $C=2$}

The above result and%
\begin{eqnarray}
X_{1,1,1,1}\left( x\right)  &=&\alpha \frac{1}{8}\left( 2+x^{2}\right)
\left( -2+\left( 2+x^{2}\right) e^{\frac{x^{2}}{2}}\Gamma \left( 0,\frac{%
x^{2}}{2}\right) \right) , \\
X_{0,0,1,1}\left( x\right)  &=&X_{1,1,0,0}\left( x\right) =\alpha \frac{1}{4}%
\left( -2+\left( 2+x^{2}\right) e^{\frac{x^{2}}{2}}\Gamma \left( 0,\frac{%
x^{2}}{2}\right) \right) , \\
X_{0,1,1,0}\left( x\right)  &=&X_{1,0,0,1}\left( x\right) =\alpha \frac{1}{2}%
\left( 1-\frac{1}{2}x^{2}e^{\frac{x^{2}}{2}}\Gamma \left( 0,\frac{x^{2}}{2}%
\right) \right) .
\end{eqnarray}

\subsection{ $C=3$}

All of the above results and%
\begin{eqnarray}
X_{0,0,2,2}\left( x\right)  &=&X_{2,2,0,0}\left( x\right) =\alpha \frac{1}{16%
}\left( -12-2x^{2}+\left( 8+8x^{2}+x^{4}\right) e^{\frac{x^{2}}{2}}\Gamma
\left( 0,\frac{x^{2}}{2}\right) \right) , \\
X_{1,1,2,2}\left( x\right)  &=&X_{2,2,1,1}\left( x\right) =\alpha \frac{1}{32%
}\left( 2+x^{2}\right) \left( -12-2x^{2}+\left( 8+8x^{2}+x^{4}\right) e^{%
\frac{x^{2}}{2}}\Gamma \left( 0,\frac{x^{2}}{2}\right) \right) ,  \notag \\
X_{2,2,2,2}\left( x\right)  &=&\alpha \frac{1}{128}\left(
8+8x^{2}+x^{4}\right) \left( -12-2x^{2}+\left( 8+8x^{2}+x^{4}\right) e^{%
\frac{x^{2}}{2}}\Gamma \left( 0,\frac{x^{2}}{2}\right) \right) ,
\end{eqnarray}%
and%
\begin{eqnarray}
X_{0,2,2,0}\left( x\right)  &=&X_{2,0,0,2}\left( x\right) =\alpha \frac{1}{16%
}\left( 4-2x^{2}+x^{4}e^{\frac{x^{2}}{2}}\Gamma \left( 0,\frac{x^{2}}{2}%
\right) \right) , \\
X_{1,2,2,1}\left( x\right)  &=&X_{2,1,1,2}\left( x\right) =\alpha \frac{1}{32%
}\left( 4+x^{2}\right) \left( 4+2x^{2}-\left( 4x^{2}+x^{4}\right) e^{\frac{%
x^{2}}{2}}\Gamma \left( 0,\frac{x^{2}}{2}\right) \right) ,
\end{eqnarray}%
and%
\begin{eqnarray}
X_{2,1,0,1}\left( x\right)  &=&X_{1,2,1,0}\left( x\right) =X_{0,1,2,1}\left(
x\right) =X_{1,0,1,2}\left( x\right)  \\
&=&\alpha \frac{1}{8\sqrt{2}}\left( 4+2x^{2}-\left( 4x^{2}+x^{4}\right) e^{%
\frac{x^{2}}{2}}\Gamma \left( 0,\frac{x^{2}}{2}\right) \right) .  \notag
\end{eqnarray}

\end{widetext}

\end{document}